\newcommand{\fcgs}{\ifmmode erg~cm^{-2}~s^{-1[B}\else 
erg~cm$^{-2}$~s$^{-1}$\fi}
\newcommand{\lcgs}{\ifmmode erg~~s^{-1[B}\else erg~s$^{-1}$\fi}
\newcommand{\fnucgs}{\ifmmode erg~cm^{-2}~s^{-1}~Hz^{-1}\else 
erg~cm$^{-2}$~s$^{-1}$~Hz$^{-1}$\fi}
\newcommand{\lnucgs}{\ifmmode erg~s^{-1}~Hz^{-1}\else 
erg~s$^{-1}$~Hz$^{-1}$\fi}
\newcommand{\kms}{\ifmmode~{\rm km~s}^{-1}\else ~km~s$^{-1}~$\fi}
\newcommand{\mone}{\ifmmode ^{-1}\else$^{-1}$\fi}
\newcommand{\mtwo}{\ifmmode ^{-2}\else$^{-2}$\fi}
\newcommand{\ax}{\ifmmode{\alpha_x} \else $\alpha_x$\fi} 
\newcommand{\aE}{\ifmmode{\alpha_E} \else $\alpha_E$\fi} 
\newcommand{\nh}{\ifmmode{\rm N_{H}} \else N$_{H}$\fi}
\newcommand{\nhgal}{\ifmmode{ N_{H}^{Gal}} \else N$_{H}^{Gal}$\fi}
\newcommand{\nhintr}{\ifmmode{ N_{H}^{intr}} \else N$_{H}^{intr}$\fi}
\newcommand{\nhtot}{\ifmmode{ N_{H}^{tot}} \else N$_{H}^{tot}$\fi}
\newcommand{\meangamma}{\ifmmode{\langle\Gamma\rangle} \else 
$\langle\Gamma\rangle$\fi}
\newcommand{\lopt}{\ifmmode l_{opt} \else $~l_{opt}$\fi}
\newcommand{\loglopt}{\ifmmode{\rm log}~l_{opt} \else log$~l_{opt}$\fi}
\newcommand{\lx}{\ifmmode l_x \else $~l_x$\fi}
\newcommand{\loglx}{\ifmmode{\rm log}~l_x \else log$~l_x$\fi}
\newcommand{\aox}{\ifmmode{\alpha_{ox}} \else $\alpha_{ox}$\fi} 
\newcommand{\auv}{\ifmmode{\alpha_{uv}} \else $\alpha_{uv}$\fi} 
\newcommand{\auvx}{\ifmmode{\alpha_{uvx}} \else $\alpha_{uvx}$\fi} 
\newcommand{\luv}{\ifmmode l_{uv} \else $~l_{uv}$\fi}
\newcommand{\logluv}{\ifmmode{\rm log}\,l_{uv} \else log$\,l_{uv}$\fi}
\newcommand{\ein}{{\em Einstein{\rm}}}
\newcommand{\ros}{{\em ROSAT{\rm}}}
\newcommand{\IUE}{{\em IUE}}

\newcommand{\lapprox}{<\atop^\sim}  
\newcommand{\ciii}{\ifmmode{{\rm C\,III]}} \else C\,III]\fi}
\newcommand{\civ}{\ifmmode{{\rm C\,IV}} \else C\,IV\fi}
\newcommand{\heii}{\ifmmode{{\rm He\,II}} \else He\,II\fi}
\newcommand{\lya}{\ifmmode{{\rm Ly}\alpha}\else Ly$\alpha$\fi}
\newcommand\lyb{\ifmmode {\rm Ly}\beta \else Ly$\beta$\fi}
\newcommand{\mgii}{\ifmmode{{\rm Mg\,II}} \else Mg\,II\fi}
\newcommand{\ovi}{\ifmmode{{\rm O\,VI}} \else O\,VI\fi}
\newcommand{\siiv}{\ifmmode{{\rm Si\,IV}} \else Si\,IV\fi}
\newcommand{\ew}{\ifmmode{W_{\lambda}} \else $W_{\lambda}$\fi}
\newcommand{\wciii}{\ifmmode{W_{\lambda}({\rm C\,III]})} \else 
$W_{\lambda}$(C\,III])\fi}
\newcommand{\wciv}{\ifmmode{W_{\lambda}({\rm C\,IV})} \else 
$W_{\lambda}$(C\,IV)\fi}
\newcommand{\wheii}{\ifmmode{W_{\lambda}({\rm He\,II})} \else 
$W_{\lambda}$(He\,II)\fi}
\newcommand{\wlya}{\ifmmode{W_{\lambda}({\rm Ly}\alpha)}\else 
$W_{\lambda}$(Ly$\alpha$)\fi}
\newcommand\wlyb{\ifmmode{ W_{\lambda}({\rm Ly}\beta )} \else 
$W_{\lambda}$(Ly$\beta$)\fi}
\newcommand{\wmgii}{\ifmmode{W_{\lambda}({\rm Mg\,II})} \else 
$W_{\lambda}$(Mg\,II)\fi}
\newcommand{\wovi}{\ifmmode{W_{\lambda}({\rm O\,VI})} \else 
$W_{\lambda}$(O\,VI)\fi}
\newcommand{\wsiiv}{\ifmmode{W_{\lambda}({\rm Si\,IV})} \else 
$W_{\lambda}$(Si\,IV)\fi}
\newcommand{\rciii}{\ifmmode{{\rm C\,III]/Ly} \alpha} \else 
C\,III]/Ly$\alpha$\fi}
\newcommand{\rciiiciv}{\ifmmode{{\rm C\,III]/C\,IV}} \else C\,III]/C\,IV\fi}
\newcommand{\rciv}{\ifmmode{{\rm C\,IV/Ly} \alpha} \else C\,IV/Ly$\alpha$\fi}
\newcommand{\rheii}{\ifmmode{{\rm He\,II/Ly} \alpha} \else 
He\,II/Ly$\alpha$\fi}
\newcommand{\rmgii}{\ifmmode{{\rm Mg\,II/Ly} \alpha} \else 
Mg\,II/Ly$\alpha$\fi}
\newcommand{\rovi}{\ifmmode{{\rm O\,VI/Ly} \alpha} \else O\,VI/Ly$\alpha$\fi}
\newcommand{\roviciv}{\ifmmode{{\rm O\,VI/C\,IV}} \else O\,VI/C\,IV\fi}
\newcommand{\rsiiv}{\ifmmode{{\rm Si\,IV/Ly} \alpha} \else 
Si\,IV/Ly$\alpha$\fi}
\documentstyle[12pt,aas2pp4]{article}
\received{December 29, 1995}
\accepted{March 14, 1996}
\tighten

\begin{document}

\title{ THE RELATIONSHIP BETWEEN THE HIGH ENERGY CONTINUUM
AND EMISSION LINES IN QSOS: A LOW-REDSHIFT SAMPLE }

\author{Paul J. Green\altaffilmark{1} }

\affil{Harvard-Smithsonian Center for
Astrophysics, 60 Garden St., Cambridge, MA 02138}
\altaffiltext{1}{pgreen@cfa.harvard.edu}

\begin{abstract}

  Photoionization models dictate that many prominent quasar emission
lines are sensitive to both the luminosity and shape of the quasars'
high energy continuum - primarily the extreme ultraviolet (EUV) and
soft X-ray continuum.  Unfortunately, the EUV band is severely
obscured by Galactic absorption.  Using data from the adjacent UV and
soft X-ray bandpasses, we initiate the first large-scale, multi-line
investigation of correlations between the QSO soft X-ray continuum and
line emission in a sample of QSOs observed by \ein~ and \IUE.

  We present a new error analysis for objective, automated line
measurements, which enables us to include the information contained in
weak or undetected lines.  We tabulate more than 300 UV emission line
equivalent widths from \IUE~ spectra of 85 QSOs in the atlas of
Lanzetta et al. (1993), then characterize the distributions of line
equivalent and velocity widths (\ew\, and FWHM).  We then compare
these line parameters to the QSO continuum spectral energy
distributions from optical through soft X-ray wavelengths, using
survival analysis to incorporate any non-detections for X-ray flux
and/or UV emission lines.  Several correlations noted in previous
studies are {\em not} reproduced here. However, we illustrate that the
exclusion of undetected lines from such studies may spuriously enhance
apparent correlations.

 We find significant correlations between \ew\, and UV luminosity
(e.g., the well-studied Baldwin effect) for \lya, \civ, \heii, and
\ciii.  \wciii\, and \wheii\, also show previously unreported
correlations with X-ray luminosity which, for \ciii, appears to be
primary. The line ratios \rciii~ and \rheii\, both show strongest
dependence on \lx.  \wlya\, correlates strongly with spectral slopes
\auv\, and \aox\, (between 2500\AA\, and 2~keV), but {\em not} with
X-ray luminosity.

  Using these results, we argue that one simple geometrical
interpretation of the Baldwin effect (BEff) as a result of a
distribution of disk inclinations is not plausible.  We also provide
evidence that the BEff weakens or disappears when the line emission is
correctly compared to the luminosity in the continuum bandpass
relevant to its production.  We thus support the interpretation of the
BEff as a change in spectral energy distribution with luminosity, and
we predict that no BEff relative to X-ray luminosity should be found
for Fe\,II or Mg\,II emission lines.  Extensions of our method to
samples of a wider redshift/luminosity range will be presented in a
later paper, which will test these predictions.

\end{abstract}


\section{INTRODUCTION}
\label{intro}


\subsection{The Ionizing Continuum of Quasars}
\label{seds}

  The majority of the nearly 8000 quasars known to date were
discovered either via their prominent optical and ultraviolet (OUV)
emission lines, or from their distinct colors in these bandpasses.
The production of emission lines in QSO spectra is widely attributed
to photoionization and heating of the emitting gas by the UV to X-ray
continuum (e.g., Ferland
\& Shields 1985, Krolik \& Kallman 1988).  Individual emission lines from
a given ion are particularly sensitive to photons of energy above the
corresponding ionization threshold.  As an example, the continuum flux
relevant to the production of Ly$\alpha$ emission is above 13.6eV,
while He\,II$\lambda$1640 is produced by photons above the 54eV
ionization edge of He$^{+}$, which at 228\AA\, is in the EUV.  Note
however that the production of many emission lines may be sensitive to
continuum energy ranges both softer and harder than the ionization
potential of the species in question because such photons may ionize
from excited states and also heat the gas via free-free and H$^-$
absorption. Many important lines respond to the extreme ultraviolet
(EUV) or soft X-ray continuum.  Unfortunately, the EUV band is
severely obscured by Galactic absorption.  However, constraints on the
EUV ionizing continuum are available both through analysis of the
emission lines, and through the adjacent UV and soft X-ray windows.

   Both radio-loud (RL) and radio-quiet (RQ) quasars are seen to have
soft ($\lapprox$1~keV) X-ray emission that exceeds the extrapolation
from the power-law continuum observed at higher energies (e.g., Turner
\& Pounds 1989, Masnou et al. 1992). This X-ray `soft excess' has often
been interpreted as the high energy continuation of the big blue bump
(BBB), possibly thermal emission from the surface of an accretion disk
(although see Barvainis 1993).  From the optical/UV side, the bump is
an upturn in emission toward shorter wavelengths commonly observed in
quasar spectral energy distributions (SEDs; e.g., Elvis et al. 1994).
Somewhere in the EUV band, the SEDs must peak and turn down again to
meet the observed X-ray emission.

\subsection{ Emission Lines as Continuum Diagnostics}
\label{balqsos}

  There are pressing reasons to investigate the relationship between
available measurements of their high energy continuum and the OUV
emission lines in QSOs.  First is to investigate observational
constraints on photoionization models for the broad line region (BLR)
of active galactic nuclei (AGN): do spectral energy distributions
(SEDs) directly determine emission line strengths or line profile
parameters?  Conversely, do similar emission line parameters in QSOs
provide empirical testimony for similar high energy SEDs?

  The overall similarity of QSO emission line spectra had been taken
as evidence of fairly uniform, robust physical conditions in the BELR,
which encouraged the assumption that clouds in the BELR inhabit a
narrow swath of parameter space (in density, size, and ionization
parameter).  Early photoionization pioneers such as Mushotsky \&
Ferland (1984) ran models on a single cloud. Refinements using cloud
ensembles showed a reduced dependence of total line emission on
intrinsic QSO SEDs (Binette et al. 1989).  Details of individual
clouds or even clouds in a single ``zone'' can be lost in the mix, and
correlations between continuum shape and observed line parameters
diluted.  Baldwin et al.  (1996) reiterate that averaging of emission
from clouds with a wide variety of properties (but uniformly large
columns) results in QSO line spectra robustly consistent with those
observed.  Correlations of \ew\, with SED would provide evidence
against such models.

  Perhaps emission lines can be used to infer the strength and shape
of the high energy SED (Krolik \& Kallman 1988, Zheng 1991), even in
the presence of extrinsic effects such as absorption along the line of
sight.  As an example, radio-quiet QSOs with broad UV absorption lines
(BALs) are now known to exhibit markedly weak X-ray emission as a
class (Green et al. 1995, Green \& Mathur 1996).  The similarity of
emission-line properties in BAL and non-BAL QSOs (Weymann et al. 1991)
has been cited as evidence that orientation is the cause of the BAL
phenomenon (i.e., {\em all} radio-quiet QSOs have BAL clouds).  If
similar emission lines indeed vouch for similar intrinsic high energy
SEDs, then the large observed \aox\, values for BAL QSOs are likely to
be caused by strong absorption along the line-of-sight rather than by
differences in their intrinsic SEDs.  However, the UV and X-ray
absorbers have yet to be positively identified as one (e.g., see the
techniques of Mathur 1994).  Since BAL QSOs {\em may} be heavily
absorbed, the question of whether similar emission lines are testimony
for similar intrinsic SEDs must be answered through study of
line/continuum correlations in unabsorbed QSOs.  The simple question
of whether line equivalent width \ew\, correlates with \aox, for
example, remains to be explored across a range of emission lines and
for QSO samples spanning a range of luminosities.

\subsection{\bf The Baldwin Effect and Changes in Continuum Shape with 
Luminosity}

  If the proportionality between line and continuum strength were
linear, then diagnostics such as line ratios and equivalent widths
would be independent of continuum luminosity.  Baldwin (1977) first
noticed that in high redshift quasars, the equivalent width
(hereafter, \ew) of the CIV $\lambda1550$\AA\, emission line in
quasars decreases with increasing UV ($1450$\AA) luminosity.  The
Baldwin effect (BEff) was also found to be strong for ions such as
OVI, NV, He\,II, CIII], Mg\,II, and Ly$\alpha$ (e.g., Tytler \& Fan
1992, Zamorani et al. 1992). Several possible explanations for the
BEff have been offered, one being a dependence of blue bump strength
on luminosity.

 The shape of the continuum (i.e., the SED) of quasars does appear to
correlate with luminosity.  In the UV regime, Zheng \& Malkan (1993)
found that the UV continuum increases in strength relative to the
optical toward higher luminosities, and that the strength of the BEff
decreases once the effect of the increasing UV (BBB) continuum is
removed. In the X-ray bandpass, the largest, most uniform study --
ROSAT All-Sky Survey (RASS) observations of 908 QSOs in the Large
Bright Quasar Survey (the LBQS/RASS; Green et al. 1995) -- confirmed
earlier reports (e.g., Wilkes et al. 1994, Tananbaum et al.  1986)
that the hypothetical power-law index between UV (rest
$\lambda2500$\AA) and soft X-ray regimes, \aox\, increases
significantly with luminosity. The increase in \aox\, is equivalent to
a {\em decrease} in soft X-ray relative to UV emission with increasing
luminosity.  There are also hints (e.g., Schartel et al. 1996) that
the soft X-ray spectral index
\ax\, of QSOs may decrease with luminosity and/or redshift.  The
decrease in \ax\, could mean that the soft X-ray spectrum {\em
hardens} with increasing redshift and/or luminosity
\footnote[1]{Redshift and 
luminosity dependence can be hard to disentangle in magnitude-limited
surveys, but several recent results, e.g., Wilkes et al. 1994, confirm
the primacy of (optical) luminosity in the correlation with \aox.}.
Alternatively, a soft excess may shift out of the \ros\, passband
toward higher redshift, and/or move toward lower energies in higher
luminosity sources.  Any or all of these trends of continuum shape
could strongly influence the efficiency of the ionizing continuum, and
should affect observed emission line strengths and
ratios. Investigations of the relationship between emission line and
continuum strengths abound in the literature, but only a handful of
small samples have been studied relating the shape and strength of the
{\em high energy} QSO continuum to emission lines.


  Zheng, Kriss, \& Davidsen (1995; hereafter ZKD) find a strong
anti-correlation between the \ew\, of OVI$\lambda1034$ and \aox.
Although some models predict such behavior for other UV emission
lines, no other such trends have been observed.  The increase in
\aox\, with luminosity in QSOs when combined with the observed BEff is
not nearly sufficient to explain the trend in \wovi\, with \aox.  What
might be responsible?  Higher luminosity QSOs may undergo spectral
evolution such that fewer photons from a soft X-ray excess/BBB
component are available for ionization.

  The intriguing results of ZKD are based on a variety of published
X-ray fluxes, and a heterogeneous compilation of rest-frame UV spectra
(2 from HUT, 16 from \IUE, 14 from $HST$, and 29 ground-based),
excluding all non-detections.  Thus, although they may well prove
robust, such results are open to challenge on the basis of the strong,
diverse selection effects inherent in such a sample.  On the other
hand, even in complete, flux-limited samples of QSOs (which often
constitute a large fraction of other more heterogeneous samples) there
is a strong correlation between redshift and luminosity.  At a {\em
given} redshift, the more luminous objects will usually have higher
signal to noise (S/N) spectra.  As a result, most of the weak-lined
QSOs remaining in a sample that ignores non-detections will be
luminous (a Malmquist bias).  In addition, noise that randomly
enhances the apparent line strength will bump low luminosity objects
into the sample with spuriously high line \ew\, (an Eddington bias).
Thus the apparent statistical significance of line/continuum
correlations may spuriously enhanced by a {\em combination} of
selection effects if undetected lines are left out of the sample.
Although some general selection effects in line/continuum studies of
the BEff have been considered in the literature (e.g., Zamorani et
al. 1992), few studies can be found incorporating line error estimates
and upper limits to line \ew, both essential to unbiased
line/continuum studies.

Here we initiate a line/continuum investigation of wide scope, using
(1) large, homogeneous samples (2) uniform data and analysis, and (3)
a wider range of lines and (consequently) ionization potentials.  We
outline new error analysis for a simple automated line measurement
technique (\S~\ref{meas}), and include limits in all analyses.  To
facilitate further study, we tabulate these data for individual
QSOs. Via correlation tests (see
\S~\ref{results} for details), we seek to determine which 
of \luv, \auv, \lx, or \aox\, dominates emission line formation, or at
least which parameter most reliably predicts measured line parameters.
In combination with other (e.g., higher redshift/luminosity) samples,
these data, techniques, and results should prove useful for further
studies of the effect of QSO SEDs on the broad emission line region
(BELR).  One such followup study is now underway, using LBQS and RASS
data (Green et al. 1996).

\section{Sample}
\label{sample}


  Multiwavelength line/continuum studies are strongly affected by
variability.  The slope and intensity of optical, UV, and soft X-ray
continua are known to vary out of proportion and out of phase to each
other, and are only occasionally correlated (Reichert et al. 1994;
Clavel et al. 1992).  Simultaneous multiwavelength coverage is very
difficult to obtain, and in any case may offer only a slight advantage
for understanding the intrinsic physics; emission lines respond to
continuum variations with time-delays that must be determined
separately for each emission line and each object (Reichert et
al. 1994, Pogge \& Peterson 1992).  To compensate for these
limitations, we prefer large samples with the most homogeneous data
and analysis, and we use averages of multiple exposures whenever
practicable.  Data sets that permit such averaging are important,
since much of the scatter in observed line/continuum correlations like
the BEff are due to variability (Kinney et al. 1990).

\smallskip
\centerline{\bf  The UV Spectra}
\label{lowzuv}

The {\em International Ultraviolet Explorer} (\IUE) satellite has
provided several thousand UV spectra of QSOs, BL Lacs, and Seyferts
since its launch.  Kinney et al. (1990) selected a subset of 69 of
these objects with 3 or more repeated observations for uniform
co-adding, using an optimized extraction technique.  Using slightly
less restrictive criteria, and similar extraction techniques,
Lanzetta, Turnshek, \& Sandoval, (1993, hereafter LTS93) compiled 260
high quality spectra.  They graciously provided these spectra,
accompanied by their 1-$\sigma$ error arrays, so that uncertainties
for all measurements can be derived.  Continuum fits were also
provided.  Details of the extraction, dereddening, and continuum
fitting techniques are described in LTS93. We exclude BL Lacs (they
generally have no emission lines) and Seyfert galaxies (to avoid
aperture effects and contamination from the host galaxy).  From the
LTS93 compilation, we have selected the subsample of all QSOs (their
class 85) with reliable redshifts, yielding 180 objects.

Both metal-line and BAL QSOs are likely to be absorbed in soft X-rays
(e.g., Green \& Mathur 1996, Mathur 1994), and there is as yet no
published soft X-ray study of other absorbed QSOs.  We therefore
remove from the sample all QSOs with absorbers as listed in Hewitt \&
Burbidge 1993 (known BALs, measured damped Ly$\alpha$, optical or UV
absorption).  We also remove 3 candidate damped Lya systems from
Lanzetta, Wolfe, \& Turnshek (1995), leaving 97 objects.  Removal of
the BAL QSO IRAS~0759+651, and a `possible' BAL 0043+039 (Tom Barlow,
private communication), leaves an \IUE~ sample of 95 QSOs. Since the
detection of absorption in any QSO spectrum is dependent on S/N, some
absorbed systems probably remain.  However, we feel the exclusion of
absorbed QSOs where possible enhances our chances of characterizing
the {\em intrinsic} emission line/continuum relationship.

Finally, we remove 2 QSOs for which none of the emission lines studied
here fall into the available \IUE~ spectra (1435$-$015 and
2216$-$038), and 8 QSOs with noisy spectra for which no emission lines
are detected (0318$-$196, 0935+417, 1006+817, 1114+445, 1215+113,
1257+346, 1352+011, and 1402+436).  In the final sample of 85 QSOs,
both short wavelength (SWP) and long wavelength (LW) spectra are
available for 62 QSOs. Only SWP spectra are available for 15, and only
LW spectra for 8 QSOs.  The redshifts of our sample range from 0.1 to
2.3, with mean and median $<z>$ of 0.55 and 0.36, respectively.

This compilation of high quality \IUE~ spectra of QSOs is necessarily
heterogeneous.  Many of the QSOs are included simply because they are
bright or peculiar in some other waveband.  Others have some unique
property which warranted their study in the UV.  Although the sample
is thus not ideal, it could be called representative, since for
objects with redshifts $z>0.05$ in the Hewitt \& Burbidge (1993,
hereafter HB93) catalog, the LTS93 \IUE~ sample contains about $85\%$
of all QSOs with $V<16$, and 50\% of those with $V<17$.  More
importantly, it is simply the largest low-redshift sample with
homogeneous UV spectra currently available.

\bigskip
\centerline{\bf  The Soft X-ray Fluxes}
\label{xrays}

The quasar database of Wilkes et al. (1994; hereafter WEA94) includes
estimates of the X-ray count rates, fluxes and luminosities for 514
QSOs and Seyfert 1 galaxies observed with the \ein~ IPC.  All objects
were previously known via radio or optical selection and most were
targets of the X-ray observations. Although like the \IUE~ sample, the
WEA94 targeted sample is heterogeneous, it again represents the
largest, most homogeneous data set currently available. By requiring
that the QSOs in the \IUE~ sample have \ein~ soft X-ray data available
in WEA94, we define the \IUE/\ein~ sample of 49 objects.

\section{Data Analysis}
\label{meas}

With spectra of lower S/N, the use of Gaussian fitting to measure line
fluxes and profiles is questionable, and tends to leave out the line
wings, which in QSOs contain a substantial fraction of the line flux.
Another common measurement of line width uses the second moment
of the flux about the mean (or median), but is severely affected by
even low-level wings.  We therefore measured line parameters from the
\IUE~ atlas spectra using a summation procedure detailed in Robertson
(1986).  We integrate line fluxes between rest-frame wavelengths listed
in Table~1, using a local linear continuum determined from bands on
either side of the emission line.  The continuum is taken as best-fit
least-squares line through the mean flux values in these bands. 
Continuum points more than 3$\sigma$ from the mean are iteratively
rejected until a maximum of 10 iterations or a minimum of 9 pixels per
band has been reached.  In the few cases where only one continuum band
is covered by the spectrum, we assume a constant continuum level fixed
at the mean value of the measurable band.  If more than 10\% of the
line region is missing, no measurement is performed.  For spectra with
adequate wavelength coverage, our measurement of the equivalent width
(\ew), full width at half maximum (FWHM) and asymmetry parameter are
described in Appendix~\ref{params} The line measurement errors are
estimated directly from the 1-$\sigma$ error spectrum when available,
and otherwise from the noise in the continuum (see
Appendix~\ref{errors}).

We set a line `detection' threshold of 5 times the errors as computed
in Appendix~\ref{errors}, and include upper limits at that level for
lines weaker than the threshold.  The largest source of systematic
error is the choice of continuum.  Our equivalent width measurements
may differ systematically from those of other studies, since our
measurements are simple and automated.  However, they should be
internally consistent, and thus most useful for correlation with
continuum properties.  We consider 6 UV emission lines
(Ly$\beta+$O\,VI, Ly$\alpha +$N\,V, Si\,IV$+$ O\,IV], C\,IV,
He\,II$+$O\,III], and Al\,III$+$C\,III).  Table~2 lists line
equivalent widths and errors for all 85 QSOs in the \IUE\, sample.
For brevity, and since they show few correlations in this sample with
SEDs, we do not present a table of the FWHM measurements.  Spectral
coverage of Mg\,II was available for only 10 QSOs, so we do not list
these data. Other lines (e.g., C\,II$\lambda$2326) are excluded
entirely since they are generally too weak or blended to detect in
spectra of rather low S/N.  The shortened names (e.g. \ovi\, rather
than Ly$\beta+$O\,VI) we use throughout the text are indicated in the
last column of Table~1.

All luminosities are calculated assuming $H_0=50$ km s\mone ~Mpc\mone
and $q_0=0.5$. To derive optical luminosities, we use the $B$
magnitudes listed in WEA94 or HB93.  If only $V$ magnitudes are
listed, we assume $B-V=0.3$.  We include a reddening correction of
$E_{B-V}={\rm max}[0,(-0.055 + 1.987\times10^{-22}\nhgal]$ and
$A_B=4~E_{B-V}$.  The Galactic neutral hydrogen column density,
\nhgal, is adopted from WEA94.  A magnitude to flux
conversion constant of 48.36 for B magnitudes (Hayes and Latham 1975)
yields for the emitted flux at 2500\AA:
$${\rm log[f_{em}(2500)] =  -19.34 + \alpha_o
log\Bigl(\frac{2500}{4400}\Bigr)  }$$ 
\begin{equation}
{ ~ ~ ~ ~ ~ ~ ~ ~ ~ ~ ~ ~ ~ - 0.4(B-A_{B}+\Delta B)}
\end{equation}
The correction $\Delta B$ includes the effects of both emission lines
and continuum slope. We derive $\Delta B$ by integrating the $B$ band
transmission function over a composite QSO spectrum (Francis et al.
1991).  We used a Matthews \& Sandage $B$ curve (FWHM=944\AA,
$\lambda_c=4460$\AA.  We provide this $k$-correction in Table~3.  For
the redshifts relevant to this study, $\Delta B$ corresponds well with
$\alpha_o=-0.23$, where $f_{\nu}\propto \nu^{\alpha_o}$.

  UV continuum flux and UV spectral slopes are determined using the
\IUE\, continuum fits of LTS93.  Second order continuum fits to
the log-log of these ($f_{\lambda}$) spectra provide the continuum slope
$a_{uv}$ and normalization $b_{uv}$.  These parameters yield the
observed flux
\begin{equation}
{\rm log}f^o_{\lambda 1450}= a_{uv}{\rm log}\lambda + b_{uv}
\end{equation}
The UV spectral slope is $\alpha_{uv}= -2 - a_{uv}$. The rest-frame
monochromatic flux at $\lambda1450$, $f_{uv}$, is then given by 
\begin{equation}
{\rm log}f_{uv}= {\rm log}f^o_{\lambda 1450} + (1+\alpha_{uv}){\rm
log}(1+z)
\end{equation}

  \ein~ broadband (0.16 - 3.5~keV) fluxes corresponding to $\ax=-0.5$
were taken from WEA94 (who prefer the convention $f_{\nu}\propto
\nu^{-\alpha_x}$).   This value of \ax~ is most appropriate for RL
QSOs.  RL QSOs on average have flatter
X-ray spectral indices \ax~ (Wilkes \& Elvis 1987; Schartel et al.
1996).  Published spectral fits to the \ein~ data are only available
for about 18 QSOs in our sample.  The assumption of a mean spectral
index $\ax=0.5$ translates (via the X-ray counts-to-flux conversion
factor and $k$-correction) into errors of $\lapprox$30\% in the
2\,keV X-ray luminosity \lx. The slope \aox\, is for a
hypothetical power law connecting rest-frame 2500\,\AA~ and 2~keV,
so that $\aox\, = 0.384~{\rm log} (\frac{ \lopt}{\lx })$.  Objects with
large \aox\, thus have stronger optical emission relative to X-ray.
We tested an analogous quantity, \auvx, defined similarly between
1450\,\AA~ and 2\,keV.  We find that \auvx\, is so tightly correlated
with \aox\, that its use as an alternative or complementary measure of
QSO SEDs is probably not warranted.  This is because the UV is nearly
100 times closer to the optical than to the soft X-ray bandpass.  Our
characterizations of continuum spectral energy distributions are
presented in Table~4.

Although more than half of the \IUE~ sample consists of radio-loud
(RL) QSOs, we do not separate these from radio-quiet (RQ) QSOs in this
study.  Except for reports of differences in \civ~ asymmetries (Corbin 
\& Francis 1994), the emission line spectra of RL and RQ QSOs are very
similar (Steidel \& Sargent 1992; Corbin 1992).  Such asymmetry
differences may also persist between core and lobe-dominant RL QSOs
(Brotherton et al. 1995). The largest problem may be variability. 
However, the \IUE~ spectra are averaged over a variety of timescales,
as are some of the \ein~ IPC fluxes.  

\section{Results}
\label{results}

\subsection{UV Emission Line Parameters, and Correlations Between Them}
\label{linecorrs}

  The sample median and mean (with its error) of emission line \ew\,
and FWHM, and of continuum parameters, are listed in Table~5. These
were determined using the survival analysis package ASURV (LaValley,
Isobe \& Feigelson 1992).  We thus incorporate the information in
non-detections, which may constitute more than 50\% of the data in some
cases (e.g., for Si\,IV$+$O\,IV).  In survival analysis, the median is
always well defined. If, however, the lowest (highest) point in the
data set is an upper (lower) limit, the mean is not well defined,
since the distribution is not normalizable, and so the outlying
censored point is redefined as a detection in our analyses.  In 
Table~5, for these cases we list the value of that redefined limit,
where the distribution is truncated. 

Our method yields EW distributions
quite similar to those of other studies (e.g., Baldwin, Wampler, \&
Gaskell 1989, Cristiani \& Vio 1990, Francis et al. 1991, Laor et al.
1995).  As a further means of comparison, we have measured the LBQS
composite spectrum of Francis et al. (1991) using our technique, and
find equivalent widths of 5.5, 49.7, 9.3, 31.0, 6.9, and 19.9 for \ovi,
\lya, \siiv, \civ, \heii, and \ciii, respectively.  Our measurements
yield \ew~ that are $0.93\pm 0.07$ times those found by Francis et al. 
(1991). From this figure we exclude a comparison of \wheii\, (our \ew\, value
is 58\% of theirs), since we use a higher continuum estimate
for this line.  Due to the existence of the BEff, we should
not expect our measured mean \ew\, measurements to be the same in the
\IUE~ sample as in other samples, unless their $<z>$ and $<\logluv>$
are similar.

  Among the emission lines, we find a strong 3-way correlation among
\wciv\, \wlya, and \wheii.  For all other
parameter pairs showing significant correlations in {\em both}
Spearman Rank and Generalized Kendall Rank tests, we display the
best-fit regressions in Table~6 whenever the number of emission
line measurements exceeds 30.  Measured FWHM for each line correlates
strongly with the 
\ew\, of that line.  We note also a significant correlation between
\rciiiciv\, and FWHM(\ovi).  However, since this is based on only 22
data points (including 13 limits), we postpone discussion to a later
paper.  We do not confirm significant correlations between \rciv~ and
\ciii/\civ~ (Mushotsky \& Ferland 1984), or between FWHM(\ciii) and
FWHM(\civ) (Corbin \& Francis 1994).

\subsection{SED Parameters, and Correlations Between Them}
\label{sedcorrs}

  As expected (given the predominance of magnitude-limited samples),
luminosities in different wavebands correlate strongly with one
another.  Distance-independent parameters such as \aox\,
generally show more scatter.  We do find a significant
anti-correlation in the \IUE/\ein\, sample of \aox\, with X-ray
luminosity ($P=0.18\%$, with slope $-0.13$). Quoted probabilities here
and in Table~6 are for the assumption of a null hypothesis (no
correlation) using a Generalized Spearman Rank (ASURV).  Similar
trends have been noted previously for low-redshift samples (e.g.,
Corbin 1993; Wang et al. 1996).  

In the \IUE/\ein\, sample, the positive correlation
between \aox\, and \loglopt\, observed in larger samples (e.g., Green
et al. 1995, WEA94, Tananbaum et al. 1986) is seen only marginally
($P<6\%$), but has both slope and intercept consistent with previous
results. The various adopted optical and X-ray $k$-corrections  make
only small differences to the derived \aox\, values.  
More importantly, this correlation is found to be weak or
possibly absent for \loglopt$\lapprox$31 (Avni et al. 1995), which is
close to the mean luminosity of our (and Corbin's) sample.  

\subsection{Correlation of UV Emission Lines with SEDs }
\label{linesedcorrs}

  We find several strong correlations ($P<2\%$) between emission line
EW and continuum SED parameters.  Comparing to luminosities, \wlya,
\wciv, \wheii, and \wciii\, all show significant anti-correlations
with \luv, i.e. a BEff.  \wheii\, and
\wciii\, also correlate strongly with \lx.  Since \luv~ and \lx~ are
very strongly correlated, these may be a secondary effect.  To test
for the primary relationship, we use the ASURV bivariate Spearman
Ranks as input to multivariate Partial Spearman Rank (PSR) analysis
(Kendall \& Stuart 1976).  \wciii\, correlates most strongly with
\lx\, ($P_{PSR}=0.059$, and a PSR of $\rho=-0.273$, while its
correlation withs.
\luv\, has $P_{PSR}=0.180$ and $\rho=-0.167$).  The primary relationship of
\wheii\, appears to be with \luv\, ($P_{PSR}=0.035$, $\rho=-0.310$ for \luv;
$P_{PSR}=0.10$, $\rho=-0.216$ for \lx).

\begin{figure} 
\plotone{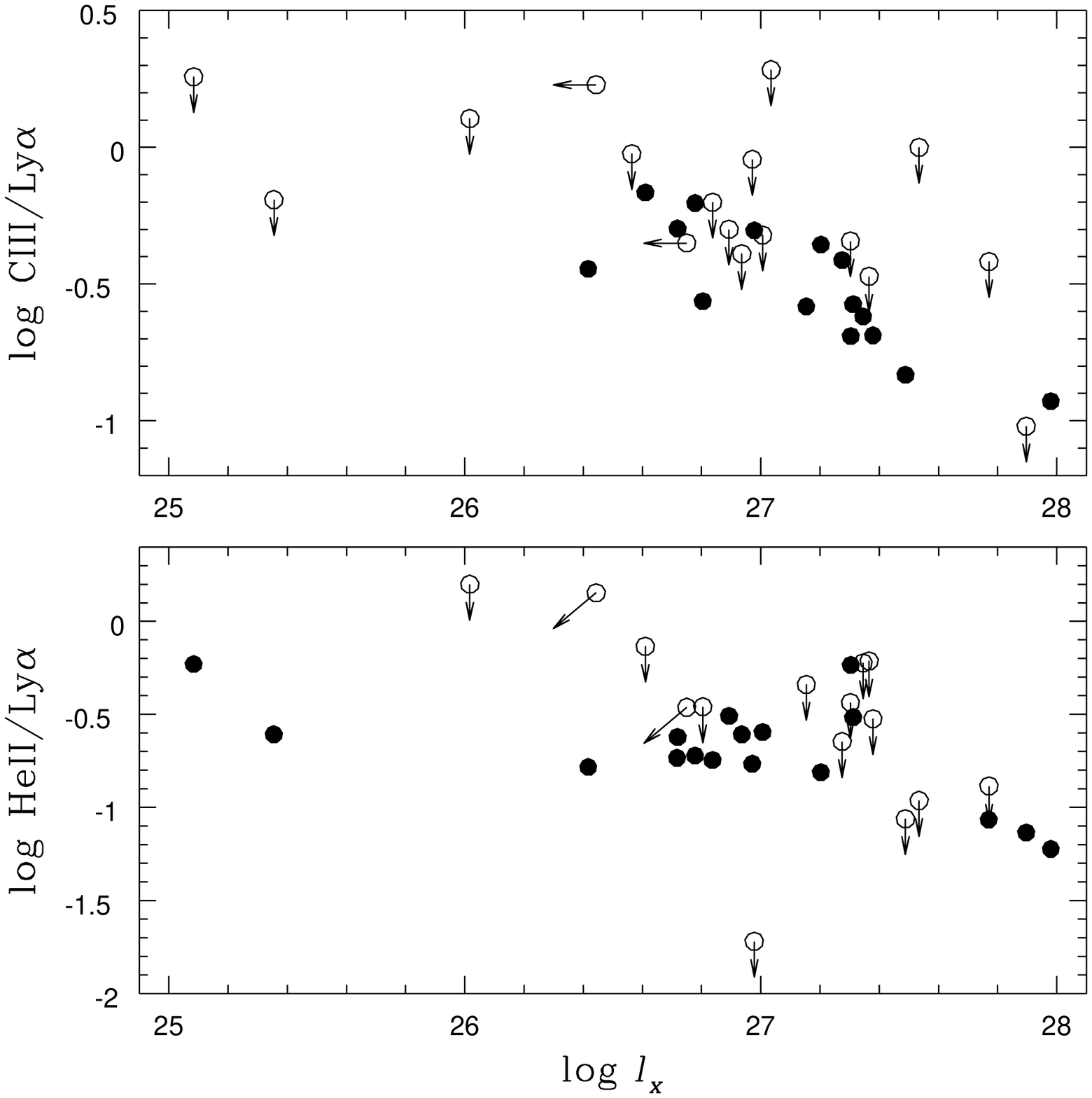}
\vfill
\caption{Line ratios (relative to \lya) vs. X-ray luminosity for
\ciii\, and \heii.   Arrows denote upper limits
to line ratios which when tilted, are X-ray upper limits as
well.  Several very high
upper limits are excluded in these plots, but have no significance
to the statistical results. Plots of \ew\, vs. X-ray luminosity appear
similar, with slightly more scatter.  
}
\label{frat_xeml}
\end{figure}

  We also test a number of line {\em ratios} against SED parameters.
The line ratio \rciii\, shows a significant correlation with X-ray
luminosity.  The ratio \rheii\, depends similarly on both \luv\, and \lx\,
(see Table~6) but appears to be most sensitive to \lx\,
($P=0.033$, $\rho=-0.322$ for \lx; $P=0.278$, $\rho=-0.105$ for \luv).
These correlations are illustrated in Figure~\ref{frat_xeml}.  Plots
of \ew\, vs. X-ray luminosity appear similar, with slightly more scatter. 

\begin{figure} 
\plotone{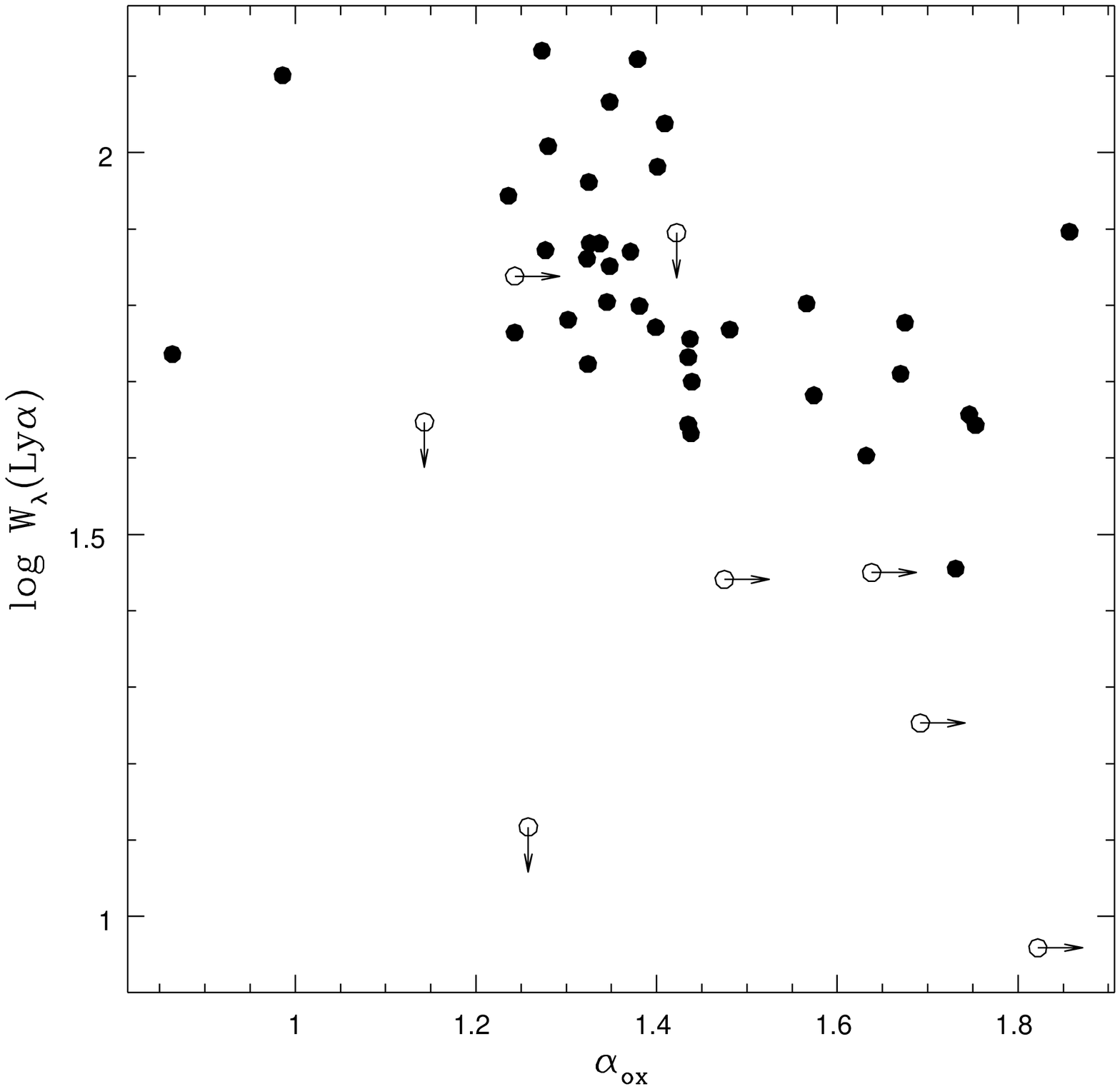}
\vfill
\caption{\lya\, equivalent width vs. \aox.  Arrows denote limits.  
We find a strong inverse correlation of \wlya\, with \aox.  Since
\aox\, is known to increase with luminosity, the anti-correlation
between \wlya\, and \aox\, shown here could be a secondary effect.
However, we do not observe an anticorrelation of \wlya\, with X-ray
luminosity, as might be expected in such a case. 
}
\label{fwlya_aox}
\end{figure}

  Testing the relationship of line \ew\, with continuum {\em shape}
parameters, we find a strong inverse correlation of \wlya\, with
\aox\, (Figure~\ref{fwlya_aox}).  Since \aox\, is known to increase
with OUV luminosity, the anti-correlation between \wlya\, and
\aox\, could be a secondary effect.  PSR tests do not reveal which of
\logluv~ or \aox\, has the primary relationship to \wlya\, 
($P_{PSR} < 0.005$ for both).  Simple substitution of the observed
\aox(\luv) relation into the observed \wlya(\luv) relation found here
produces a significantly steeper slope ($\sim -1$) than we derive here
for \wlya(\aox).  We suspect that this is a mathematical artifact of
line regressions through a population with high dispersion, and
with error in both variables.  That an 
\ew\, to \aox\, correlation appears for \lya\, but not other lines
that show a BEff could be a result of the higher S/N of the
\lya\, line.  We suspect that the \wlya(\aox) correlation is primary  
however, since \wlya\, does not correlate with X-ray luminosity \lx,
and the latter clearly bears a stronger relationship to \luv\, than
does \aox.  We also point out that the trend in FWHM here could
be strongly affected by changes in N\,V, or the \lya/N\,V ratio.

  We find few correlations between our measured line FWHM and SED
parameters.  However, FWHM(\lya) correlates strongly to \luv\, and to
\auv.  Since \auv\, and \luv\, were from independent fits to the
entire spectrum (LTS93), we expect that these trends are not an
artifact of our choice of continuum used for line integration, which
uses bands only redward of \lya\, (Table~1).  However, in
data of such low S/N (the mean is about 8 for the \IUE~ sample
spectra), the FWHM measurements are not very robust.  Asymmetry
measurements, which require still higher moments of the flux
distribution, are even less so, and are thus excluded from
consideration here.  We will construct composite spectra from
subsamples with similar SEDs in a later paper to increase the overall
S/N of these tests.  This will be of particular use in incorporating
the LBQS/RASS sample, for which X-ray flux stacking will permit an
acceptable detection fraction (Green et al. 1995).

\section{Discussion}
\label{discuss}

\subsection{Confirmations \& New Results}
\label{new}

  We confirm significant correlations between \ew\, and UV luminosity
(e.g., the well-studied Baldwin effect) for \lya, \civ, \heii, and
\ciii.  Models of optically thick, geometrically thin accretion disks
(Netzer 1987) have been successful in explaining several of these
line/continuum correlations.  Limb darkening and projected surface area
effects in these models call for a UV continuum flux that is strongest
face-on, and highly anisotropic compared to the harder
(e.g., soft X-ray) ionizing flux.  Quasar disks, particularly the area
emitting the UV flux, are presumably much smaller than the BLR. Thus a
random selection of objects, differing only in disk inclination,
results in measurements of constant line 
luminosities but yields UV continuum luminosities that vary with
aspect, thereby producing the observed anticorrelation between \ew\,
and \luv.  We note that if indeed X-ray emission is more isotropic,
much weaker correlations would be observed between \ew\, and \lx.  In
addition, disks viewed face-on (smaller \ew) would also appear to have
boosted UV (larger \aox).  All of these predictions hold true in the
current study for \lya.  However, within this geometric model, we
would certainly expect to see the same set of correlations with \civ,
yet there is no corresponding decrease of \wciv\, with \aox.  

We report for the first time significant anti-correlations between
\wciii, \wheii, and X-ray luminosity which, for \ciii, appears to be
primary.  This correlation also seems to contradict the simplest
geometrical explanation of the BEff.  However, neither side of the
argument may be physically meaningful unless we can
contrast line \ew\, to the continuum relevant to the line's
production.  For instance, 
\ciii\, line emission should depend almost exclusively on the Lyman
continuum (between about 13 and 25eV), which is at least an order of
magnitude in energy below the soft X-ray bandpass sampled here. \heii\,
line emission is spurred by continuum photons between 54 and 150eV.  UV
lines such as \lya\, and \civ\, on the other hand, should also depend
on energies (from $300 - 400$eV) closer to the \ein\, bandpass (see
Table~4 of Krolik \& Kallman 1988; hereafter KK88).  

  Not only photons that serve to ionize the species in question
contribute to emission line production for that species.  Ionization
from excited states and heating via free-free and H$^-$ absorption
also help determine the line's principal ionizing/heating continuum
(PIHC).  Traditionally, the BEff is seen when comparing UV emission
line \ew\, to UV luminosity.  For some cases in the current study we
are able to compare the line \ew\, somewhat more directly to a portion
of its PIHC using extant soft X-ray observations.  For these cases,
\lya\, and \civ, the {\em X-ray} BEff is not significant.  For \heii\,
and \ciii\, lines, where the entire PIHC is softer than the \ein\,
bandpass (KK88), we show that a significant BEff persists relative to
soft X-ray luminosities.  This would suggest that the BEff weakens
substantially when the continuum luminosity used for comparison to
\ew\, is  close to the PIHC of the line.  Changes in the relative
normalization of the `near-line' continuum to the PIHC thus enhance
the traditional BEff.  This adds weight to previous arguments
(beginning with Malkan \& Sargent 1982) that the BEff is caused by
changes in continuum shape.  The increase in luminosity may primarily
be due to an increase in UV/EUV/soft X-ray emission (the BBB, possibly
the thermal signature of an accretion disk) over the underlying
power-law continuum.  It has been proposed that the BBB shifts toward
lower energies at higher (OUV) luminosities This would entail (1) a
strong increase in \luv; (2) a weaker increase in
\lx; and (3) an increase in \aox.  The response of line flux and \ew\,
depends in a fairly complicated manner on the relative peak energies
of the BBB and the PIHC, and the BBB normalization relative to the
power-law continuum.  However, given adequate multiwavelength
coverage, it is possible in principle to contrast the theoretical and
observed line response.

  In this picture, a stronger BEff might be expected for species of
higher PIHC.  Since traditionally, the BEff contrasts \ew\, to the
luminosity near the line, the slope or normalization of the BEff may
be accentuated if the actual PIHC is more distant in energy from that
luminosity.  There is indeed evidence for such a trend (Zheng et
al. 1992; Espey, Lanzetta, \& Turnshek 1993).

  Another prediction of this picture is that lines whose PIHC is entirely
observable will show a much weaker BEff relative to that
continuum.  Good candidates for such lines would have PIHCs very near
the soft X-ray bandpass, with no contribution (as with \lya\, and \civ)
from softer continuum components.  Some examples are observationally
challenging: He\,I$\lambda$5876, with PIHC $300 - 500$eV, is
quite weak; C\,II$\lambda$326, with PIHC above 800eV, is in the EUV;
O\,I$\lambda$8447 (PIHC $>$600eV) requires spectra at least into the
near-IR for most QSOs.  Space-based detection of Ne~VIII$\lambda$774,
which has an ionization energy of 207eV, is difficult as well (Hamann
et al. 1995). Study of other lines is more tractable: Fe\,II
lines in the UV have their PIHC above 500eV; Fe\,II lines in the
optical are principally due to continuum above 800eV. There is
some preliminary  vindicating evidence that optical ($\lambda4570$)
Fe\,II emission is intimately linked to its observable PIHC: Shastri et
al. (1993) and Laor et al. (1994) found that QSOs with strong optical
Fe\,II emission show softer (steeper) X-ray spectral slopes.  Green
et al. (1995) found that QSOs in the LBQS with strong UV Fe\,II
emission (based on 
the iron feature under [Ne~IV]$\lambda 2423$) are anomalously X-ray
bright in the \ros\, passband.  To eschew any nascent complacency, we
point out that Corbin (1993) did find an anti-correlation between
\ew(Fe\,II) and soft X-ray luminosity.  Previous photoionization models
may be insufficient in the case of Fe\,II, which a) could partly arise
from collisional excitation in regions optically thick to X-rays
(e.g., Kwan \& Krolik 1981) and b) probably need drastically revised
photoionization cross-sections (Bautista \& Pradhan 1995).  

  Mg\,II$\lambda$2798 appears to be the best candidate for testing our
predictions.  This line has a PIHC between 600-800eV, and is much more
easily and objectively measured than are the broad Fe\,II multiplets.
Studies of Mg\,II emission lines vs. X-ray continuum appear to have been
overlooked.  Though an insufficient number of \IUE\, spectra in this
study include the Mg\,II region, we are currently pursuing such
correlations with the LBQS/RASS sample (Green et al. 1996).
 
  The line ratios \rciii~ and \rheii\, both show their strongest
dependence on \lx.  KK88 state that, since pure recombination is a poor
approximation for \lya\, in broad-line clouds, the \rheii\, line ratio
can trace only very large spectral contrasts in the continuum.  In
fact, neither of these lines shows a strong correlation across a wide
range of \aox.  By contrast, Boroson \& Green (1992) found an
anti-correlation between He\,II$\lambda 4686$/H$\beta$ and \aox.
This may indicate that no simple relation exists between
the optical and UV helium line strengths.  Boroson \& Green
had pointed out the potential utility such a relation might hold for
luminosity calibration and measurements of $q_0$.

\subsection{Non-confirmations of Previous Results}
\label{nonconf}

\begin{figure} 
\plotone{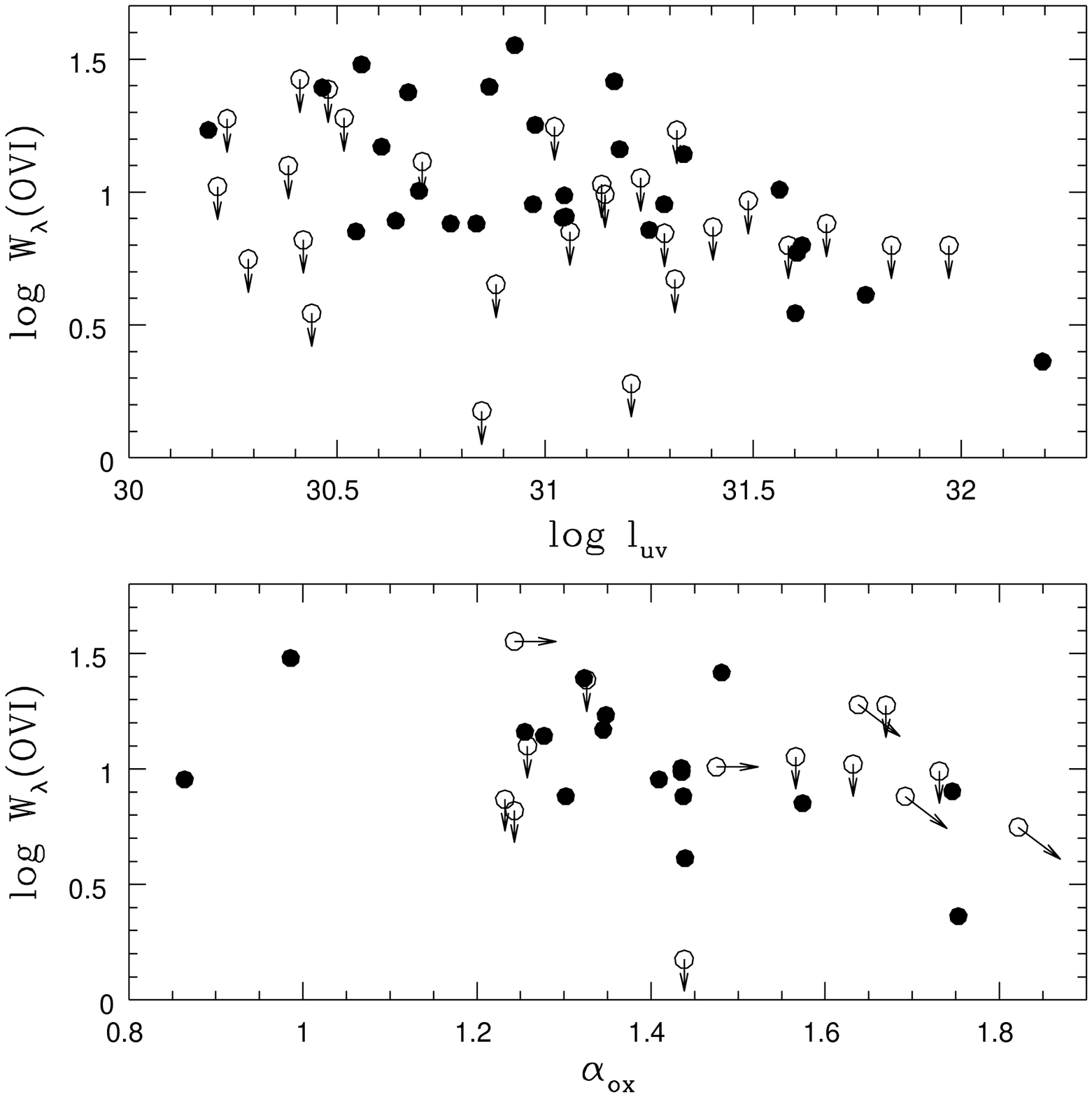}
\vfill
\caption{O\,VI] equivalent width vs. UV luminosity and \aox.
Arrows denote limits. Contrary to Zheng, Kriss, \& Davidsen (1995;
ZKD), we do not confirm 
significant ($P<2\%$) correlations between \wovi\, and either \aox\,
or \logluv\, from our  \IUE\, sample. However, when we analyze {\em
detections only}, we find significant correlations, with slopes
consistent with ZKD ($-0.81\pm0.30$ and $-0.41\pm0.08$, respectively). 
This illustrates that the exclusion of undetected lines from
line/continuum studies may spuriously enhance apparent correlations.  
}
\label{fwovi}
\end{figure}

  We do {\em not} confirm the correlation of \rovi\, with \aox\,
reported by ZKD.  They also find strong anti-correlations between
\wovi\, and both \aox\, (slope $-0.81\pm0.27$) and \logluv\, (slope
$-0.30\pm0.03$).  Using our \IUE\, sample, and including upper limits,
we cannot confirm any of these correlations at our adopted significance
level (see Figure~\ref{fwovi}).  However, when we analyze {\em
detections only},  we find significant correlations, with slopes
consistent with ZKD ($-0.81\pm0.30$ and $-0.41\pm0.08$, respectively). 
These results highlight that {\em upper limits should  always be
included to avoid spuriously significant line/continuum correlations.}
However, the ZKD result is still valid - they appear to have
detected \ovi\, for every QSO in their sample {\em a posteriori}
(e.g., the sample was not selected for strong \ovi\, emission).  
Some of the correlations seen by ZKD and others are probably
indeed intrinsic to QSOs, but are not reproduced here because of the
smaller luminosity range of our sample.  This will be tested in an
upcoming paper that 
includes a larger luminosity range, while still including limits. 
(However, to avoid contamination of the \ovi\, line region by \lya\,
forest absorption, we will still be limited to relatively low
redshifts.  Future inclusion of $HST$ data is clearly warranted, but
beyond the scope of the current study.) We note also that 27 of 32 of
the QSOs in the ZKD sample are radio-loud, for which the BEff
may be enhanced (perhaps a bias introduced by their greater variability;
Murdoch 1983).  Zamorani et al. (1992) discuss these issues further,
and find flatter slopes in \wciv\, vs. \luv\, using
optically-selected QSOs than were found in the PKS sample of Baldwin et
al. (1989).  The slope we find here is intermediate.

The early single cloud models of Mushotsky \& Ferland (1984) first
suggested that the observed increase of \aox\, with luminosity could
cause the BEff, and would also predict, if anything, a {\em
negative} correlation of \rciv\, with luminosity. Baldwin, Wampler, \&
Gaskell (1989; BWG hereafter) and Kinney et al. (1987, 1990) found
that both \rciv\, and \rciiiciv\, are inversely correlated with UV
luminosity.  We confirm neither of these correlations.  
The same single cloud models also predicted that \lya, \civ, and
\ciii\, line ratios should be relatively independent of both \aox\, and
$\alpha_x$ because these lines should not originate from X-ray-heated
zones deeper in the emission-line clouds.  Here we find instead that
\wlya\, does depend an \aox, and that \rciii\, anti-correlates with
X-ray luminosity. 

  Optically thin clouds, which may become fully ionized in hydrogen,
(Goad et al. 1993, Shields et al. 1995) may also demand inclusion in
models of the BLR.  The response of line emission to continuum 
changes may be flat or even negative in optically thin clouds, which
can can help reproduce 1) the observed difference in the lag of the
high- and low-ionization lines relative to the continuum in Seyferts
(e.g., Reichert et al. 1994); 2) some ultraviolet absorption features
and 'warm absorber' behavior in the X-ray regime (e.g., Mathur 1994);
3) the intrinsic BEff.  The latter is a strong decrease in
line equivalent width that occurs as the luminosity increases
in individual variable Seyferts (Kinney et al. 1990).  The global BEff
has a much shallower slope, which might be explained by a decrease in
the covering factor $f_c$ of the optically thin component, due to more
efficient outflow of thin clouds in intrinsically brighter sources.

  We briefly explore reasons why several correlations between line and
continuum parameters that have been found significant in other studies
are not reproduced here.  (1) Some studies, while more heterogeneous, have
embraced a wider luminosity range than our \IUE~ sample, often
including optical data and higher redshift QSOs. Several results
(e.g., Wampler et al. 1984, Kinney et al. 1987) have shown that for
QSOs of lower luminosity (\logluv$\lapprox$31), the \ew -luminosity
relations flatten or disappear.  The mean \logluv\, of the
\IUE\, sample we investigate here is 30.6, with only about a third of
the sample having \logluv$>31$.  Even at higher luminosities, a very
large range of luminosity is often required to overcome intrinsic
scatter in the global Baldwin relation, and scatter induced by
variability (e.g., Kinney et al. 1990).  (2) The exclusion of
undetected emission lines from many previous samples spuriously
enhances the apparent statistical significance of line/continuum
correlations.  We note that a slight relaxation of our significance
criterion from $P<2\%$ to $P$$\lapprox$5\% would have retrieved
several previously reported correlations.  (3)  Although the
\IUE/\ein\, sample is large, and the data and analysis homogeneous, 
the QSOs therein were selected for observation by these satellites
for a variety of reasons, so this sample cannot be considered
truly complete or homogeneous.

\section{Conclusions}
\label{conclude}

  Complex activity is likely to be associated with the nuclear
environment in QSOs.  Rapid star formation and evolution, supernovae,
accretion/merging of galaxies or protogalactic fragments, and a
supermassive accreting  black hole may all contribute.  Although QSO
spectra are surprisingly homogeneous given such a flamboyant cast, the
simplest geometric and photoionization models do not succeed in
explaining the relationship of QSO emission lines to the observed
continuum.  This may partly be alleviated if the line emission can be
directly compared to its principal ionizing/heating continuum.  

  Objective, automated line measurements including line upper limits
are crucial to avoid spurious enhancement of apparent line/continuum
correlations.  We find significant correlations between \ew\, and UV
luminosity (e.g., the well-studied Baldwin effect) for \lya, \civ,
\heii, and \ciii.  \wciii\, and \wheii\, also show previously unreported
correlations with X-ray luminosity which, for \ciii, appears to be
primary. The line ratios \rciii\, and \rheii\, both show strongest
dependence on \lx.  \wlya\, correlates strongly with spectral slopes
\auv\, and \aox\, (between 2500\AA\, and 2~keV), but {\em not} with
X-ray luminosity.

  Using these results, we argue that one simple geometrical
interpretation of the BEff that assumes ionizing X-ray emission to be
more isotropic than UV continuum emission is not plausible.  If indeed
the BEff were a result of a distribution of disk inclinations in this
case, weak anti-correlations of line \ew\, with X-ray luminosity would
be expected at best.  The significant anti-correlations of \ciii\, and
\heii\, emission with \lx\, thus render the simplest geometrical
model unlikely.

When we are able to compare the line \ew\, most
directly to a portion of its PIHC using extant soft X-ray 
observations (for \lya\, and \civ) the {\em X-ray}
BEff is not significant.  For \heii\, and \ciii\, lines,
where the entire PIHC is softer than the \ein\, bandpass (KK88), 
a significant BEff persists relative to soft X-ray
luminosities.  We thus argue that the
BEff weakens or disappears when the line emission is compared
to the luminosity in the bandpass of its principle photoionizing
continuum.   This supports an interpretation of
the BEff as a change in spectral energy distribution with
luminosity.  We predict that no BEff relative to soft X-ray
luminosity should be found for Fe\,II or Mg\,II emission lines. 
Extensions of our method to samples of a wider redshift/luminosity
range would test these predictions.   

Now that we have outlined a technique for the efficient measurement of
large numbers of comparatively low S/N QSO spectra, we will apply it
to the largest, most uniformly-selected such sample to date, the
LBQS. Optical spectra and X-ray fluxes or upper limits are available
for 908 QSOs in the LBQS from the {\em ROSAT} All-Sky Survey.
Analysis of that database will be combined with the results presented
here to offer a truly wide luminosity baseline for further study of
the interdependence of QSO continuum and emission line properties.

  Of course, we would have preferred for each line an accurate measure
of its rest-frame principle ionizing/heating continuum.  For \ovi,
this means the the `He\,I' continuum ($24.5-54.4$eV). For \lya\,
\ciii\, and \civ\, this means the ranges $13.6-24.5$eV (the `Lyman
continuum') dominates (KK88). \ciii\, and
\civ\, should depend also on ionizing photons from $0.3-0.4$keV.  Since
these include the EUV range, which is observationally inaccessible for
all but a handful of nearby AGN (Marshall et al. 1995), we have
attempted an indirect examination of the relevant continuum via the
adjacent UV and X-ray luminosities, and through \aox.  The true
strength of the EUV is best estimated using spectral index and
normalization in the adjacent UV and soft X-ray bands together.
Slopes are available in the UV data set, but the great majority of
X-ray data provide only net counts in the \ein\, bandpass ($\sim
0.16-3.5$keV).  Unfortunately, there are too few QSOs in the \IUE~
sample with published \ax\, (about 18, judging from Elvis et al. 1994)
to permit any convincing statistical tests.  The 2keV monochromatic
fluxes used in \aox\, are thus derived from these \ein\, counts
assuming a single power-law slope, and absorption due to Galactic
$N_H$ only.  Although there may be absorption (either warm or cold;
see e.g., Netzer 1993) intrinsic to the QSOs, this effect is unlikely
to be strong.  A small soft excess above the power-law, which could be
important to the ionizing continuum is, however, expected in many of
the QSOs (Fiore et al.  1994).  Estimates of X-ray spectral slopes in
the \ros\, band ($0.1-2.4$keV) for more than 100 bright QSOs should be
available from ROSAT observations within the next few years (e.g.,
Bade et al.  1995).  These data should prove a valuable addition to
the studies initiated here.

Hearty thanks to Ken Lanzetta for providing his \IUE~ spectral atlas,
including error spectra and continuum fits, in digital form.  Paul
Eskridge helped simplify the task of PSR analysis. Craig Foltz
provided a copy of the LBQS composite spectrum. I gratefully
acknowledge Avi Loeb for our discussions of the emission line error
analysis.

\clearpage

\appendix

\section{Definitions of the Line Parameters}
\label{params}

Following Robertson (1986), the line equivalent width, $W_{\lambda}$, is
\begin{equation}
\label{eqeqw}
        W_{\lambda} = \sum_{i=\lambda_1}^{\lambda_2}
        \Bigl(\frac{f_{l,i}}{f_{c,i}} - 1 \Bigr) \Delta\lambda_i,
\end{equation}
\noindent where $f_{l,i}$ and $f_{c,i}$ represent the flux
in the line and continuum, respectively at the $i$th pixel.

The FWHM, $\delta$  is taken to be
\begin{equation}
\label{eqfwhm}
        {\rm FWHM}_{\lambda} = 1.1775(\Lambda_H - \Lambda_L).
\end{equation}
\noindent Here, $\Lambda_H$ and $\Lambda_L$ are the wavelengths where
16\% and 84\% of the line flux, respectively, are reached while
integrating over the line region, corresponding to the $\pm1\sigma$
points for a (noise-free) Gaussian line profile. These wavelengths
are defined by the following equations:
\begin{displaymath}
\label{eqlambdal}
        \sum_{-\infty}^{\Lambda_L}
        \Bigl(\frac{f_{l,i}}{f_{c,i}} - 1 \Bigr) \Delta\lambda
        = 0.1587~ W_{\lambda}
\end{displaymath}

\noindent and

\begin{displaymath}
\label{eqlambdah}
        \sum_{-\infty}^{\Lambda_H}
        \Bigl(\frac{f_{l,i}}{f_{c,i}} - 1 \Bigr) \Delta\lambda
        = 0.8413~ W_{\lambda}.
\end{displaymath}

The definition of the asymmetry parameter, $\xi$, is
\begin{equation}
\label{eqasymm}
        \xi = \frac{100}{\delta} \left[ 0.5 \left(
\Lambda_H + \Lambda_L \right) - \Lambda \right]
\end{equation}
\noindent Here, $\Lambda$ is the median wavelength,  where
50\% of the line flux is reached while integrating over the
line region:
\begin{displaymath}
\label{eqlambda}
        \sum_{-\infty}^{\Lambda}
        \Bigl(\frac{f_{l,i}}{f_{c,i}} - 1 \Bigr) \Delta\lambda
        = 0.5~ W_{\lambda}
\end{displaymath}

\section{Error Computations for Line Parameters}
\label{errors}

\subsection{Errors on the Wavelengths $\Lambda, \Lambda_L, \Lambda_H$}

  The error in $\Lambda$ is
\begin{displaymath}
\sigma_{\Lambda}^2= \left( \frac{\partial \Lambda}{\partial W_{\lambda}}
\right) ^2 \sigma_{W_{\lambda}}^2 \approx  
\frac{ \sigma_{W_{\lambda}}^2 }{ 
\left( \frac{dW_{\lambda}} {d\Lambda} \right)^2 
}
\end{displaymath}

Recalling that if 
\begin{displaymath}
        y= \int_{-\infty}^{a}f(x)~dx
\end{displaymath}
\noindent then $ \frac{\partial y}{\partial a} = f(a) $,
and so
\begin{equation}
   \frac{ dW_{\lambda}} {d\Lambda}= \frac{1}{0.5}
\left( \frac{f_l - f_c}{f_c} 
\right)_{\Lambda}.
\end{equation} 
Thus, 
\begin{displaymath}
\sigma_{\Lambda}^2 =
\left( \frac{0.5 f_c}{f_l - f_c}\right)_{\Lambda}^2 
\sigma_{W_{\lambda}}^2
\end{displaymath} 
Similarly,
\begin{minipage}[t]{2in}
\begin{displaymath}
\sigma_{\Lambda_L}^2 =
\left( \frac{0.1587 f_c}{f_l - f_c}\right)_{\Lambda_L}^2 
\sigma_{W_{\lambda}}^2, 
\end{displaymath} 
\end{minipage}

\noindent and

\begin{minipage}[t]{2in}
\begin{displaymath}
  ~ \sigma_{\Lambda_H}^2 =
\left( \frac{0.8413 f_c}{f_l - f_c}\right)_{\Lambda_H}^2
\sigma_{W_{\lambda}}^2.
\end{displaymath} 
\end{minipage}

\subsection{Errors on Line Parameters $W_{\lambda},~ \delta,$ and $\xi$ }

Propagation of errors from eq.~\ref{eqeqw} yields, for a single
pixel in the line,
\begin{displaymath}
        \sigma^2_{ W_{\lambda},i } = 
           \Bigl[  \Bigl(\frac{1}{f_{c,i}}\Bigr)^2
           \sigma_{l,i}^2 \Delta 
                + \Bigl(\frac{f_{l,i}}{f^2_{c,i}}\Bigr)^2
           \sigma_{c,i}^2 \Bigr] \Delta \lambda_i ^2
\end{displaymath}
Since our continuum level is fit from a large number of points, its
formal error should be low.  We nevertheless conservatively assume
that the error in the continuum across the line region $\sigma_{c,i}$
is equal to the error in the signal itself $\sigma_{l,i}$.  Thus, we
estimate the variance in the equivalent width to be
\begin{equation}
\label{eqsigmaeqw}
        \sigma^2_{W_{\lambda}} = \sum_{i=\lambda_1}^{\lambda_2}
           \Bigl[ \Bigl(\frac{1}{f_{c,i}}\Bigr)^2
                + \Bigl(\frac{f_{l,i}}{f^2_{c,i}}\Bigr)^2
           \Bigr] \sigma_{l,i}^2 (\Delta \lambda_i )^2
\end{equation}

When $\sigma_{l}$ is not directly available from a 1-$\sigma$ error
spectrum, we estimate it from the mean variance
$\overline{\sigma^2_c}$ in the continuum bands as $\sigma^2_{l,i}=
\overline{ \sigma^2_{c}} (\frac{f_{l,i}}{f_{c,i}})$.  Tests show that
this substitution yields error estimates generally within about 30\%
of those obtained using the actual 1-$\sigma$ error spectra.

 Propagation of errors for eq.~\ref{eqfwhm} yields
a variance in FWHM of 
\begin{displaymath}
        \sigma_{\delta}^2 = 1.1775^2 (\sigma_{\Lambda_H}^2 +
\sigma_{\Lambda_L}^2),
\end{displaymath}

 For the variance in the asymmetry parameter,
propagation of errors for eq.~\ref{eqasymm} yields
\begin{displaymath}
        \sigma_{\xi}^2 = \frac{1}{\delta ^2}
\left[ \left(1.1775^2 \xi ^2 + 50^2 \right)
 (\sigma_{\Lambda_H}^2 + \sigma_{\Lambda_L}^2)
+ 100^2 \sigma_{\Lambda}^2 \right]
\end{displaymath}

\clearpage
\setcounter{page}{27}

\onecolumn

\begin{references}
\reference{} Avni, Y., Worrall, D. M., \& Morgan, W. A. Jr. 1995, ApJ,
454, 673 
\reference{} Bade, N. Fink, H.H., Engels, D., Voges, W., Hagen,
H.-J., Wisotzki, L., \&  Reimers, D.N. 1995, A\&AS, 110, 469
\reference{} Baldwin, J.A. 1977,  ApJ, 214, 679.
\reference{} Baldwin, J.A., Wampler, E. J., \& Gaskell, C. M. 1989, ApJ,
338, 630 (BWG)
\reference{} Baldwin, J. A., Ferland, G., Korista, K., \& Verner, D.
1996, ApJ, in press
\reference{} Barvainis, R. 1993, ApJ, 412, 513
\reference{} Bautista, M. A., \& Pradhan, A. K. 1995,
At. Mol. Opt. Phys., 28, L173 
\reference{} Binette, L., Prieto, A., Szusziewicz, E.,
\& Zheng, W. 1989, ApJ., 343, 135
\reference{} Boroson, T. A., \& Green, R. F. 1992, ApJS, 80, 109
\reference{} Brotherton, M. S.,  Wills, B. J.,  Steidel,
C. C.. \&;  Sargent, W. L. W. 1994, ApJ, 423, 131
\reference{} Clavel, J., et al. 1992, ApJ, 393, 113
\reference{} Corbin, M. R. 1992, ApJ, 391, 577
\reference{} Corbin, M. R. 1993, ApJ, 403, L9
\reference{} Corbin, M. R., \& Francis, P. J. 1994, AJ, 108, 2016
\reference{} Cristiani, S. \& Vio, R. 1990, A\&A, 227, 385
\reference{} Elvis, M., et al. 1994, ApJSup, 95, 1
\reference{} Espey, B. R., Lanzetta, K. M., \& Turnshek, D. A. 1993,
BAAS, 25, 1448
\reference{} Ferland, G. F., \& Shields, G. A. 1985 in {\it Astrophysics of
 Active Galaxies \& Quasi-Stellar Objects,} ed. J. Miller (Mill Valley,
 Ca: University Science Books), p.57
\reference{} Fiore, F., Elvis, M., McDowell, J. C., Siemiginowska, A.,
\& Wilkes, B. J. 1994, ApJ, 431, 515 
\reference{} Francis, P. J.,  Hewett, P., Foltz, C., Chaffee, F., Weymann,
R., \& Morris, S. 1991, ApJ, 373, 465
\reference{} Goad, M. R., O'Brien, P. T., Gondhalekar, P. M.
1993, MNRAS, 263, 149
\reference{}  Green, P. J., Schartel, N., Anderson, S. F., Hewett, P. C.,
Foltz, C. B., Fink, H., Brinkmann, W., Tr\"umper, J., \& Margon, B.
1995, ApJ, 450, 51
\reference{} Green, P. J. \& Mathur, S. 1996, ApJ, 462, in press
\reference{} Green, P. J., et al., 1996, in preparation
\reference{} Hamann, F., Shields, J. C., Ferland, G. J., \& Korista,
K. T. 1995, ApJ, 454, 688
\reference{} Hayes, D. S.,  \& Latham, D. W. 1975,  ApJ 197, 599
\reference{} Hewitt, A. \& Burbidge G., 1993, ApJS, 87, 51 (HB93)
\reference{} Kendall, M., \& Stuart, A. 1976, The Advanced Theory of
Statistics, Vol. II (New York: Macmillan).
\reference{} Kinney, A. L., Huggins, P. J., Glassgold, A. E., \& 
Bregman, J. N. 1987,  ApJ, 314, 145
\reference{} Kinney, A. L., Rivolo, A. R., \& Koratkar, A. P. 1990,
ApJ, 357, 338
\reference{} Krolik, J. H. \& Kallman, T. R. 1988, ApJ, 324, 714 (KK88)
\reference{} Kwan, J. \& Krolik, J. H. 1981,    ApJ, 250, 478.
\reference{} Laor, A., Fiore, F., Elvis, M., Wilkes, B. J., \&
McDowell, J. 1994, ApJ, 435, 611
\reference{} Laor, A., Bahcall, J. N., Jannuzi, B. T.,
Schneider, D. P., Green, R. F. 1995, ApJS, 99, 1
\reference{} Lanzetta, K. M., Turnshek, D. A., \& Sandoval, J. 1993,
ApJS, 84, 109 (LTS93)
\reference{} Lanzetta, K. M., Wolfe, A. M., \& Turnshek, D. A. 1995,
ApJ, 440, 435  
\reference{} LaValley, M., Isobe, T. \& Feigelson, E.D. 1992, 
in Astronomical Data Analysis Software \& Systems, ed. D. Worrall et
al. (San Francisco: ASP)
\reference{} Malkan, M. \& Sargent, W. L. W. 1982, ApJ, 254, 22
\reference{} Marshall, H. L., Fruscione, A., \& Carone, T. E. 1995,
ApJ, 439, 90
\reference{} Masnou, J.-L., Wilkes, B. J., Elvis, M., Arnaud, K. A. \&
McDowell, J. C. 1992, A\&Ap, 253, 35
\reference{} Mathur, S. 1994, ApJ, 431, L75
\reference{} Murdoch, H. S. 1983, MNRAS, 202, 987
\reference{} Mushotzky, R. F., \& Ferland, G. J. 1984,  ApJ, 278, 558.
\reference{} Netzer, H. 1987,  MNRAS, 225, 55.
\reference{} Netzer, H. 1993,  ApJ, 411, 594
\reference{} Pogge, R. W., \& Peterson, B. M., 1992, AJ, 103, 1084
\reference{} Reichert, G. A., et al. 1994, ApJ, 42, 582
\reference{} Robertson, J. G., 1986, PASP, 98, 1220
\reference{} Schartel, N., Green, P. J., Anderson, S. F., Hewett, P. C.,
Fink, H., Brinkmann, W., Tr\"umper, J., Margon, B., \& Foltz, C. B.
1996, MNRAS, submitted
\reference{} Shastri, P.,  Wilkes, B. J.,  Elvis, M.,  \& McDowell, J.
1993, ApJ, 410, 29
\reference{} Shields, J. C., Ferland, G. J., \& Peterson, B. M. 1995,
ApJ, 441, 507
\reference{} Steidel, C. C., \& Sargent, W. L. W. 1991, ApJ, 382, 433
\reference{} Tananbaum, H., Avni, Y., Green, R. F., Schmidt, M., \&
Zamorani, G. 1986,  ApJ, 305, 57
\reference{} Turner, T. J. \& Pounds, K. A. 1989, MNRAS, 240, 833
\reference{} Tytler, D. \& Fan, X.-M. 1992, ApJS, 79, 1
\reference{} Wampler, E. J., Gaskell, C. M., Burke, W. L., \&
Baldwin, J. A. 1984, ApJ, 276, 403
\reference{} Weymann, R. J., et al. 1991, ApJ, 373, 23 
\reference{} Wilkes, B. J., \& Elvis, M., 1987,  ApJ, 323, 243
\reference{} Wilkes, B. J., Tananbaum, H., Worrall, D. M., Avni, Y.,
Oey, M. S. \& Flanagan, J. 1994, ApJS, 92, 53 (WEA94)
\reference{} Zamorani, G., Marano, B., Mignoli, M., Zitelli, V. \& Boyle,
B. J. 1992, MNRAS, 256, 238
\reference{} Zheng, W. 1991, ApJ, 382, L55
\reference{} Zheng, W., Fang, L. Z.,  \& Binette, L. 1992, ApJ, 392, 74
\reference{} Zheng, W. \& Malkan, M. A., 1993, ApJ, 415, 517
\reference{} Zheng, W., Kriss, G. A. \& Davidsen, A. F. 1995, ApJ,
440, 606 (ZKD)
\end{references}
\end{document}


\label{tlinedefs}
\begin{centering}
\begin{tabular}{llccccclc} 
\multicolumn{9}{c}{ TABLE 1 } \\
\multicolumn{9}{c}{ LINE AND CONTINUUM DEFINITIONS } \\
\hline\hline  
\multicolumn{9}{c}{ } \\
\multicolumn{2}{c}{Line}  &
\multicolumn{4}{c}{Integration Limits} & 
& Name$^a$ & Notes \\ 
Identification & Wavelength & Line & Continuum  & Continuum  & Continuum  & Continuum  &  \\
  & ~~~(\AA) & (\AA) & (\AA) & (\AA) & (\AA) & (\AA) & & \\
\hline

Ly$\beta$ + O\,VI & 1025.7 \& 1033.8 & 1018 -- 1054 & ~975 -- 1018
& 1056 -- 1148 &\ldots  & \ldots & \ovi & 1 \\
Ly$\alpha$ + N\,V & 1215.7 \& 1240.1 & 1186 -- 1280 & 1180 -- 1190 &
1280 -- 1290 & 1320 -- 1330 & 1450 -- 1480  & \lya & \\
Si\,IV + O\,IV] & 1396.7 \& 1402.3 & 1363 -- 1443 & 1320 -- 1330 &
1450 -- 1480 & 1680 -- 1700  & \ldots & \siiv & \\
C\,IV & 1549.1 & 1475 -- 1600 & 1450 -- 1480 & 1680 -- 1700 &\ldots  &
\ldots & \ldots & \\
He\,II + O\,III]  & 1640.5 \& 1664.2 & 1602 -- 1680 &  1450 --
1480 & 1680 -- 1700 &\ldots  & \ldots & \heii & 2,3 \\ 
Al\,III + C\,III] & 1857.4 \& 1908.7 & 1830 -- 1950 & 1780 -- 1830 & 
1970 -- 2030 &\ldots  & \ldots & \ciii & 3 \\
\hline
\end{tabular}\\[0.5ex]

\hskip0.4in \parbox[b]{6.5in}{
$^a$~Shorter name used throughout the text.

NOTES.--- (1) Both line and continuum region susceptible to \lya\,
forest contamination; (2) Probable contamination from \civ;
(3) Possible contamination from Fe\,II.
}
\hoffset0.0in
\end{centering}
\clearpage

\hoffset-0.5in
\label{teqw}
\begin{centering}
\begin{tabular}{lrrrrrrrrrrrr} 
\multicolumn{13}{c}{ TABLE 2 } \\
\multicolumn{13}{c}{ UV EMISSION LINE EQUIVALENT WIDTHS$^a$ } \\
\hline\hline  
\multicolumn{13}{c}{} \\
Name  & \lyb~ & err &\lya~ & err & \siiv~ & err & \civ & err & \heii & err &
  Al\,III~ & err \\
    & $+$\,\ovi &  & $+$\,NV &  & $+$\,OIV] &  &  &  & $+$\,O\,III] &  &
 $+$\,\ciii &  \\ 
\hline
0026$+$129  &   \ldots & \ldots &    ~~~  59.0 &    1.8 &    $<$   6.1 & \ldots &    ~~~  32.7 &    1.0 &    ~~~  18.3 &    0.6 &    $<$  29.6 & \ldots \\ 
0044$+$030  &    $<$  11.3 & \ldots &    ~~~  63.4 &    6.4 &    ~~~  50.3 &    5.5 &    ~~~  50.8 &    1.8 &    ~~~   9.8 &    0.6 &    ~~~  27.9 &    1.5 \\ 
0052$+$251  &   \ldots & \ldots &    ~~~  91.4 &    2.6 &    ~~~   8.0 &    1.6 &    ~~~ 119.1 &    2.5 &    ~~~  15.7 &    0.9 &    $<$  82.5 & \ldots \\ 
0117$+$213$^{~l}$  &    ~~~  10.2 &    1.4 &    ~~~  27.6 &    1.5 &   \ldots & \ldots &   \ldots & \ldots &   \ldots & \ldots &   \ldots & \ldots \\ 
0119$-$286  &   \ldots & \ldots &    ~~~  89.8 &    4.3 &    ~~~  33.5 &    2.8 &    ~~~  76.3 &    2.7 &    ~~~  12.2 &    0.9 &    $<$  36.0 & \ldots \\ 
0134$+$329  &    $<$  12.6 & \ldots &    $<$  13.1 & \ldots &    ~~~  13.9 &    2.0 &    $<$ 133.3 & \ldots &    $<$  35.4 & \ldots &    ~~~  19.3 &    2.6 \\ 
0205$+$024  &   \ldots & \ldots &    ~~~  52.8 &    4.4 &    $<$   5.1 & \ldots &    ~~~  38.5 &    1.2 &    ~~~   9.5 &    0.6 &    $<$  33.2 & \ldots \\ 
0312$-$770  &    ~~~  24.7 &    1.1 &    ~~~  72.6 &    2.5 &    ~~~   5.3 &    0.9 &    ~~~  69.7 &    1.1 &    $<$ 563.5 & \ldots &    $<$ 139.5 & \ldots \\ 
0405$-$123  &    ~~~   4.1 &    0.3 &    ~~~  50.1 &    0.8 &    $<$   6.9 & \ldots &    ~~~  25.7 &    1.1 &    ~~~   3.0 &    0.5 &    ~~~   5.9 &    0.7 \\ 
0414$-$060  &    ~~~   9.0 &    0.5 &    ~~~ 109.1 &    6.9 &    ~~~  12.9 &    1.8 &    ~~~  57.9 &    1.9 &    ~~~   9.4 &    1.4 &   \ldots & \ldots \\ 
0558$-$504  &   \ldots & \ldots &    ~~~  27.8 &    0.8 &    $<$   2.7 & \ldots &    ~~~  23.1 &    0.5 &    ~~~  10.2 &    0.3 &    $<$  14.9 & \ldots \\ 
0624$+$691  &    ~~~   7.2 &    0.6 &    ~~~  60.9 &    1.4 &    $<$   1.5 & \ldots &    ~~~  51.3 &    6.5 &    $<$  16.1 & \ldots &    ~~~  12.2 &    0.9 \\ 
0736$+$017  &   \ldots & \ldots &    ~~~ 132.5 &   10.1 &    $<$  14.5 & \ldots &    ~~~  94.5 &    3.2 &    $<$ 681.9 & \ldots &    $<$ 125.8 & \ldots \\ 
0743$-$673$^{~l}$  &    ~~~   6.3 &    1.2 &    ~~~  57.2 &    3.1 &   \ldots & \ldots &   \ldots & \ldots &   \ldots & \ldots &   \ldots & \ldots \\ 
0804$+$761  &   \ldots & \ldots &    ~~~ 101.8 &    2.8 &    ~~~  28.0 &    1.1 &    ~~~  81.7 &    1.6 &    ~~~  19.3 &    0.5 &    ~~~  63.7 &    4.2 \\ 
0837$-$120  &   \ldots & \ldots &    ~~~  87.6 &    5.5 &    $<$  13.9 & \ldots &    ~~~  91.4 &    2.5 &    $<$  32.1 & \ldots &    $<$  39.8 & \ldots \\ 
0859$-$140$^{~l}$  &    $<$ 193.0 & \ldots &    $<$  78.5 & \ldots &   \ldots & \ldots &   \ldots & \ldots &   \ldots & \ldots &   \ldots & \ldots \\ 
0906$+$484  &   \ldots & \ldots &    ~~~  59.9 &    3.0 &    $<$  11.1 & \ldots &    ~~~  52.4 &    2.2 &    ~~~  14.8 &    1.0 &    $<$  38.6 & \ldots \\ 
0953$+$414$^{~s}$ &    ~~~  24.9 &    1.3 &    ~~~  63.8 &    2.8 &    ~~~   8.9 &    0.9 &    ~~~  53.6 &    1.1 &   \ldots & \ldots &   \ldots & \ldots \\ 
1001$+$291  &    $<$   4.5 & \ldots &    ~~~  34.1 &    2.3 &    ~~~   5.6 &    0.7 &    $<$  14.4 & \ldots &    $<$  14.1 & \ldots &    ~~~  20.1 &    2.1 \\ 
1004$+$130  &    $<$   5.6 & \ldots &    ~~~   9.1 &    1.5 &    ~~~  12.0 &    1.0 &    $<$ 100.2 & \ldots &    $<$1939.9 & \ldots &    $<$  69.1 & \ldots \\ 
1007$+$417  &    $<$   7.0 & \ldots &    ~~~  79.1 &    4.9 &    $<$  21.8 & \ldots &    ~~~  72.5 &    2.8 &    ~~~   8.3 &    1.0 &    $<$   6.2 & \ldots \\ 
1008$+$133$^{~l}$  &    $<$   7.6 & \ldots &    ~~~  17.9 &    1.7 &    ~~~ 180.7 &   31.3 &   \ldots & \ldots &   \ldots & \ldots &   \ldots & \ldots \\ 
1011$-$282$^{~s}$ &    ~~~  17.1 &    3.2 &    ~~~ 116.5 &   12.7 &    $<$  11.7 & \ldots &   \ldots & \ldots &   \ldots & \ldots &   \ldots & \ldots \\ 
1012$+$008$^{~s}$ &   \ldots & \ldots &    ~~~  72.9 &    4.2 &    $<$   9.8 & \ldots &    ~~~  13.3 &    1.4 &    $<$   3.7 & \ldots &   \ldots & \ldots \\ 
1048$-$090  &    $<$  24.4 & \ldots &    ~~~  76.1 &   13.0 &    ~~~  62.9 &    7.1 &    $<$  85.0 & \ldots &    $<$  45.5 & \ldots &    ~~~  18.3 &    3.4 \\ 
1049$-$005  &    $<$  26.6 & \ldots &    ~~~  83.9 &   10.0 &    $<$   8.1 & \ldots &    $<$ 124.9 & \ldots &    $<$  78.4 & \ldots &    ~~~  18.6 &    3.5 \\ 
1100$+$772  &    ~~~   7.6 &    0.6 &    ~~~  60.4 &    1.5 &    ~~~   3.5 &    0.5 &    ~~~  89.9 &    5.6 &    $<$  18.1 & \ldots &    ~~~  12.4 &    2.0 \\ 
1103$-$006  &    $<$  13.0 & \ldots &    ~~~  37.3 &    2.4 &    $<$  12.9 & \ldots &    ~~~  27.7 &    3.9 &    $<$  10.2 & \ldots &    ~~~  20.7 &    1.6 \\ 
1104$+$167$^{~l}$  &   \ldots & \ldots &    ~~~  83.4 &   12.2 &    $<$  20.5 & \ldots &    ~~~  85.1 &    3.9 &    ~~~  13.8 &    1.1 &    $<$  11.5 & \ldots \\ 
1116$+$215  &   \ldots & \ldots &    ~~~  62.9 &    1.3 &    $<$   2.6 & \ldots &    ~~~  36.5 &    0.5 &    $<$   1.2 & \ldots &    ~~~  31.2 &    2.8 \\ 
1127$-$145$^{~s}$ &   \ldots & \ldots &   \ldots & \ldots &   \ldots & \ldots &   \ldots & \ldots &   \ldots & \ldots &   \ldots & \ldots \\ 
1136$-$135  &    $<$   4.7 & \ldots &    ~~~  63.5 &    1.6 &    $<$  17.6 & \ldots &    ~~~  50.7 &    3.1 &    ~~~   6.7 &    1.0 &    ~~~  20.0 &    1.0 \\ 
1137$+$660  &    ~~~  13.9 &    1.0 &    ~~~  74.4 &    7.9 &    ~~~  41.4 &    7.2 &    ~~~  45.9 &    3.9 &    $<$   9.7 & \ldots &    $<$  28.4 & \ldots \\ 
1146$-$037  &    $<$ 463.7 & \ldots &    $<$  44.4 & \ldots &    $<$  43.2 & \ldots &    ~~~ 268.1 &   33.5 &    $<$  24.5 & \ldots &    ~~~  51.8 &    5.0 \\ 
1148$+$549  &    $<$   6.3 & \ldots &    ~~~  54.6 &    2.3 &    ~~~  11.3 &    1.2 &    ~~~  18.0 &    1.6 &    $<$  22.9 & \ldots &   \ldots & \ldots \\ 
1151$+$117  &   \ldots & \ldots &    ~~~ 112.0 &    7.2 &    $<$  12.7 & \ldots &    ~~~  69.6 &    2.6 &    ~~~  21.8 &    1.5 &    $<$  36.6 & \ldots \\ 
1156$+$295  &    $<$   1.9 & \ldots &    $<$  15.6 & \ldots &    ~~~   8.8 &    1.6 &    $<$   5.7 & \ldots &    $<$   3.9 & \ldots &   \ldots & \ldots \\ 
1156$+$631$^{~l}$  &   \ldots & \ldots &    $<$ 133.4 & \ldots &    $<$  28.8 & \ldots &    ~~~  91.3 &    7.8 &    $<$  11.2 & \ldots &    $<$  14.3 & \ldots \\ 
1202$+$281  &   \ldots & \ldots &    ~~~  74.2 &    1.3 &    $<$   3.5 & \ldots &    ~~~ 129.1 &    1.2 &    ~~~  18.9 &    0.4 &    $<$  35.4 & \ldots \\ 
1206$+$459  &    $<$  17.6 & \ldots &    ~~~  44.5 &    2.1 &    $<$  10.5 & \ldots &   \ldots & \ldots &   \ldots & \ldots &   \ldots & \ldots \\ 
1216$+$069  &    ~~~  10.1 &    1.7 &    ~~~  53.9 &    3.9 &    ~~~   9.3 &    1.4 &    ~~~  65.0 &   10.3 &    $<$  24.7 & \ldots &    ~~~  14.1 &    2.2 \\ 
1241$+$176  &    ~~~   8.0 &    0.9 &    ~~~  45.4 &    0.5 &    $<$   6.3 & \ldots &   \ldots & \ldots &   \ldots & \ldots &   \ldots & \ldots \\ 
1248$+$401  &    $<$  17.1 & \ldots &    ~~~  39.6 &    2.5 &    $<$   8.7 & \ldots &    $<$  35.3 & \ldots &   \ldots & \ldots &   \ldots & \ldots \\ 
1259$+$593  &    $<$   7.1 & \ldots &    ~~~  26.8 &    1.1 &    $<$  27.7 & \ldots &    ~~~  58.6 &    8.6 &    ~~~  12.2 &    2.2 &    ~~~  16.6 &    1.0 \\ 
1302$-$102  &    $<$   1.5 & \ldots &    ~~~  42.9 &    0.8 &    ~~~   4.4 &    0.3 &    ~~~  15.9 &    1.8 &    $<$   9.7 & \ldots &    ~~~  16.6 &    1.0 \\ 
1307$+$085  &   \ldots & \ldots &    ~~~  76.1 &    3.4 &    $<$   8.1 & \ldots &    ~~~  93.7 &    1.9 &    ~~~  18.8 &    0.8 &    $<$  31.1 & \ldots \\ 
1309$+$355$^{~s}$ &   \ldots & \ldots &    ~~~  46.7 &    2.3 &    $<$   4.9 & \ldots &    ~~~  32.9 &    1.0 &    $<$   2.6 & \ldots &   \ldots & \ldots \\ 
\hline
\end{tabular}

\clearpage
\begin{tabular}{lrrrrrrrrrrrr} 
\multicolumn{13}{c}{ TABLE 2 - {\em Continued}} \\
\hline\hline  
\multicolumn{13}{c}{ } \\
Name  & \lyb~ & err &\lya~ & err & \siiv~ & err & \civ & err & \heii & err &
  Al\,III~ & err \\
    & $+$\,\ovi &  & $+$\,NV &  & $+$\,OIV] &  &  &  & $+$\,O\,III] &  &
 $+$\,\ciii &  \\ 
\hline
1317$+$277  &    ~~~   5.9 &    1.1 &    ~~~  27.0 &    1.0 &    ~~~   3.0 &    0.4 &    ~~~  26.5 &    1.2 &   \ldots & \ldots &   \ldots & \ldots \\ 
1318$-$113$^{~s}$ &   \ldots & \ldots &   \ldots & \ldots &   \ldots & \ldots &   \ldots & \ldots &   \ldots & \ldots &   \ldots & \ldots \\ 
1333$+$176  &    $<$   9.8 & \ldots &    ~~~  28.5 &    2.2 &    $<$  83.6 & \ldots &    ~~~  42.2 &    6.4 &    $<$   9.9 & \ldots &    ~~~   7.8 &    1.5 \\ 
1334$+$246  &   \ldots & \ldots &    ~~~  54.4 &    5.3 &    $<$  25.2 & \ldots &    ~~~  63.4 &    6.6 &    ~~~  39.7 &    2.3 &    $<$ 133.2 & \ldots \\ 
1338$+$416  &    $<$  10.7 & \ldots &    ~~~  25.8 &    1.0 &    ~~~  26.4 &    1.8 &   \ldots & \ldots &   \ldots & \ldots &   \ldots & \ldots \\ 
1352$+$183  &   \ldots & \ldots &    ~~~  71.0 &    3.2 &    $<$   7.0 & \ldots &    ~~~  85.5 &    2.3 &    ~~~  17.0 &    1.0 &    ~~~  35.8 &    6.5 \\ 
1354$+$195$^{~s}$ &    ~~~  17.9 &    1.5 &   \ldots & \ldots &   \ldots & \ldots &   \ldots & \ldots &   \ldots & \ldots &   \ldots & \ldots \\ 
1355$-$416$^{~s}$ &    ~~~  14.8 &    2.1 &    ~~~  63.7 &    3.7 &    ~~~  11.7 &    1.1 &   \ldots & \ldots &   \ldots & \ldots &   \ldots & \ldots \\ 
1356$+$581$^{~s}$ &   \ldots & \ldots &   \ldots & \ldots &   \ldots & \ldots &   \ldots & \ldots &   \ldots & \ldots &   \ldots & \ldots \\ 
1415$+$451$^{~s}$ &   \ldots & \ldots &    ~~~ 117.0 &    7.0 &    ~~~  42.2 &    3.7 &    ~~~  80.3 &    3.4 &    ~~~  17.9 &    1.1 &   \ldots & \ldots \\ 
1425$+$267  &    $<$  18.9 & \ldots &    ~~~  51.3 &    3.4 &    ~~~  40.5 &    2.2 &    ~~~ 131.3 &   19.7 &    $<$  37.7 & \ldots &    ~~~  35.1 &    3.8 \\ 
1444$+$407$^{~s}$ &    $<$   3.5 & \ldots &    ~~~  51.3 &    2.2 &    ~~~  14.0 &    0.9 &   \ldots & \ldots &   \ldots & \ldots &   \ldots & \ldots \\ 
1512$+$370  &    ~~~   7.6 &    0.5 &    ~~~  57.0 &    0.8 &    ~~~   3.5 &    0.3 &    ~~~  83.7 &    3.6 &    ~~~  17.4 &    1.7 &    ~~~  15.2 &    0.7 \\ 
1522$+$101  &    $<$   6.3 & \ldots &    ~~~  41.2 &    1.6 &   \ldots & \ldots &   \ldots & \ldots &   \ldots & \ldots &   \ldots & \ldots \\ 
1525$+$227  &    $<$  10.5 & \ldots &    ~~~  40.1 &    6.4 &    $<$   7.8 & \ldots &    $<$  32.7 & \ldots &    $<$  63.4 & \ldots &    $<$  51.2 & \ldots \\ 
1526$+$285  &    ~~~  23.7 &    4.7 &    ~~~ 120.9 &    6.3 &    $<$ 182.4 & \ldots &    $<$ 316.5 & \ldots &    $<$  99.3 & \ldots &    $<$  37.5 & \ldots \\ 
1538$+$477  &    ~~~   8.1 &    0.8 &    ~~~  75.8 &   10.0 &    ~~~  19.8 &    2.7 &    ~~~  41.7 &    1.7 &    ~~~  11.5 &    1.1 &   \ldots & \ldots \\ 
1545$+$210  &    $<$   6.6 & \ldots &    ~~~  58.1 &    3.6 &    $<$   7.4 & \ldots &    ~~~  79.9 &    5.1 &    $<$  35.6 & \ldots &    $<$  19.6 & \ldots \\ 
1552$+$085  &   \ldots & \ldots &    ~~~  78.7 &    9.0 &    $<$  20.4 & \ldots &    ~~~ 210.3 &   22.1 &    ~~~  46.3 &    4.0 &    $<$ 142.4 & \ldots \\ 
1612$+$261$^{~s}$ &   \ldots & \ldots &    ~~~ 135.7 &   11.1 &    ~~~  22.0 &    4.4 &    ~~~ 148.7 &    6.4 &    ~~~  25.1 &    1.7 &   \ldots & \ldots \\ 
1617$+$175$^{~l}$  &   \ldots & \ldots &   \ldots & \ldots &   \ldots & \ldots &   \ldots & \ldots &   \ldots & \ldots &    ~~~  41.2 &    7.6 \\ 
1630$+$377  &    ~~~   3.5 &    0.6 &    ~~~  39.5 &    1.0 &   \ldots & \ldots &   \ldots & \ldots &   \ldots & \ldots &   \ldots & \ldots \\ 
1634$+$706  &    ~~~   2.3 &    0.2 &    ~~~  44.0 &    0.4 &   \ldots & \ldots &   \ldots & \ldots &   \ldots & \ldots &   \ldots & \ldots \\ 
1641$+$399  &    ~~~   9.0 &    1.0 &    ~~~  54.5 &    3.5 &    $<$  14.5 & \ldots &    ~~~  33.8 &    1.8 &    ~~~   4.0 &    0.7 &    $<$   5.2 & \ldots \\ 
1656$+$053$^{~s}$ &    $<$   7.4 & \ldots &   \ldots & \ldots &   \ldots & \ldots &   \ldots & \ldots &   \ldots & \ldots &   \ldots & \ldots \\ 
1718$+$481  &    $<$   6.3 & \ldots &    ~~~  33.9 &    0.6 &    ~~~   5.4 &    0.9 &   \ldots & \ldots &   \ldots & \ldots &   \ldots & \ldots \\ 
1721$+$343  &    ~~~  30.2 &    1.2 &    ~~~ 126.3 &    4.2 &    $<$   3.6 & \ldots &    ~~~  94.0 &    1.4 &    $<$  11.0 & \ldots &    ~~~  18.6 &    3.0 \\ 
1803$+$676  &   \ldots & \ldots &    ~~~  95.8 &    3.7 &    ~~~   9.6 &    0.8 &    ~~~ 133.3 &    1.3 &    ~~~  15.8 &    0.4 &    ~~~  34.4 &    4.6 \\ 
1850$-$782$^{~l}$  &   \ldots & \ldots &   \ldots & \ldots &   \ldots & \ldots &   \ldots & \ldots &   \ldots & \ldots &    $<$ 140.0 & \ldots \\ 
1928$+$738  &    ~~~   7.8 &    1.3 &    ~~~  71.3 &    3.4 &    ~~~   9.1 &    1.1 &    $<$  76.0 & \ldots &    $<$  85.9 & \ldots &    ~~~  22.7 &    3.7 \\ 
2112$+$059  &    ~~~  35.7 &    6.5 &    ~~~  68.9 &    2.8 &    $<$  60.6 & \ldots &    $<$ 136.0 & \ldots &    $<$  23.8 & \ldots &    ~~~  30.8 &    1.7 \\ 
2141$+$175$^{~s}$ &    ~~~   7.1 &    1.1 &    ~~~  48.1 &    2.5 &    ~~~   7.3 &    1.1 &    ~~~  18.2 &    0.9 &   \ldots & \ldots &   \ldots & \ldots \\ 
2201$+$315  &    ~~~   9.7 &    0.8 &    ~~~  44.1 &    1.5 &    ~~~  16.1 &    0.6 &    ~~~  45.9 &    3.8 &    ~~~  25.7 &    4.1 &    ~~~   9.0 &    1.7 \\ 
2251$+$113  &    $<$  19.0 & \ldots &    ~~~  28.2 &    2.8 &    $<$   5.2 & \ldots &    ~~~  79.8 &    9.3 &    $<$  40.1 & \ldots &    ~~~  47.8 &    3.8 \\ 
2251$+$158$^{~s}$ &    ~~~  14.5 &    2.6 &   \ldots & \ldots &   \ldots & \ldots &   \ldots & \ldots &   \ldots & \ldots &   \ldots & \ldots \\ 
2302$+$029  &    $<$   9.3 & \ldots &    ~~~  12.6 &    1.7 &    $<$   6.0 & \ldots &    $<$  33.9 & \ldots &   \ldots & \ldots &   \ldots & \ldots \\ 
2344$+$092  &    ~~~  26.1 &    1.9 &    ~~~  58.6 &    5.7 &    $<$  17.6 & \ldots &    ~~~  32.3 &    2.2 &    $<$   6.4 & \ldots &    $<$  58.6 & \ldots \\
\hline
\end{tabular}\\[0.5ex]

\parbox[b]{7in}{
\hskip0.4in $^a$~Equivalent widths expressed in \AA, in the rest-frame.

\hskip0.4in $^l$~Only LW spectra available.

\hskip0.4in $^s$~Only SWP spectra available.

}

\end{centering}
\clearpage

\label{tkcorr}
\begin{centering}
\begin{tabular}{cccccccccc} 
\multicolumn{10}{c}{ TABLE 3 } \\
\multicolumn{10}{c}{ADOPTED $B$ MAGNITUDE $k$-CORRECTION$^a$ } \\
\hline\hline  
\multicolumn{10}{c}{} \\
$z$ &  $\Delta B$ & $z$ &  $\Delta B$ & $z$ &  $\Delta B$ & $z$ &  $\Delta B$ & $z$ &  $\Delta B$ \\
\hline
0.0 & ~0.000 & 	0.7 & $-$0.511 & 	1.4 & $-$0.733 & 	2.1 & $-$0.832 & 	2.8 & $-$0.724 \\ 
0.1 & $-$0.071 & 	0.8 & $-$0.529 & 	1.5 & $-$0.772 & 	2.2 & $-$0.870 & 	2.9 & $-$0.669 \\ 
0.2 & $-$0.171 & 	0.9 & $-$0.545 & 	1.6 & $-$0.811 & 	2.3 & $-$0.894 & 	3.0 & $-$0.607 \\ 
0.3 & $-$0.259 & 	1.0 & $-$0.574 & 	1.7 & $-$0.822 & 	2.4 & $-$0.879 & 	3.1 & $-$0.541 \\ 
0.4 & $-$0.349 & 	1.1 & $-$0.631 & 	1.8 & $-$0.828 & 	2.5 & $-$0.849 & 	3.2 & $-$0.466 \\ 
0.5 & $-$0.421 & 	1.2 & $-$0.671 & 	1.9 & $-$0.827 & 	2.6 & $-$0.813 & 	3.3 & $-$0.378 \\ 
0.6 & $-$0.475 & 	1.3 & $-$0.703 & 	2.0 & $-$0.823 & 	2.7 & $-$0.772 & 	3.4 & $-$0.278 \\ 
\hline
\end{tabular}\\[0.5ex]

\parbox[b]{7in}{
\hskip1.4in $^a$~From LBQS composite spectrum, for a Mathews \&
Sandage $B$ filter.
}

\end{centering}
\clearpage

\label{tsed}
\begin{centering}
\begin{tabular}{lclrrrrrrrr} 
\multicolumn{11}{c}{ TABLE 4 } \\
\multicolumn{11}{c}{ CONTINUUM ENERGY DISTRIBUTIONS } \\
\hline\hline  
\multicolumn{11}{c}{} \\
Name & $z$ &  $B$ & $F_x^{~a}$ & $N_H^{~b}$ & \auv$^{~c}$ &
$b_{uv}^{~d}$  & log~$l_{opt}^{~e}$ & log~$l_{x}^{~f}$ &
log~$l_{uv}^{~g}$ & \aox  \\  
\hline
0026$+$129 & 0.142 & 14.95  &  905 &  4.6 &  $-$0.7 & $-$11.5 & 30.54 &  26.89 & 30.12 &  1.40    \\ 
0044$+$030 & 0.624 & 15.97  &   97 &  2.9 &  $-$0.5 & $-$12.5 & 31.28 &  27.20 & 31.09 &  1.57   \\ 
0052$+$251 & 0.155 & 15.42  &  910 &  4.6 &  $-$1.4 &  $-$9.2 & 30.42 &  26.97 & 30.27 &  1.33    \\ 
0117$+$213 & 1.493 & 16.05  &  $<$  170 &  4.8 &  $-$0.2 & $-$13.9 & 32.03 &  $<$ 28.19 & 31.25 & $>$ 1.48   \\ 
0119$-$286 & 0.117 & 14.89  &  \ldots &  1.6 &  $-$1.3 &  $-$9.5 & 30.34 &  \ldots & 30.15 &  \ldots  \\ 
0134$+$329 & 0.367 & 16.62  &  410 &  4.6 &  $-$1.1 & $-$11.0 & 30.65 &  27.37 & 30.30 &  1.26    \\ 
0205$+$024 & 0.156 & 15.67  &  660 &  3.3 &  $-$0.8 & $-$11.4 & 30.29 &  26.84 & 30.05 &  1.32   \\ 
0312$-$770 & 0.223 & 16.26  &  510 &  7.5 &  $-$1.2 & $-$10.1 & 30.48 &  27.03 & 30.42 &  1.32    \\ 
0405$-$123 & 0.574 & 14.75  &  685 &  3.7 &  $-$1.1 &  $-$9.9 & 31.73 &  27.98 & 31.64 &  1.44    \\ 
0414$-$060 & 0.781 & 16.24  &  230 &  5.1 &  $-$0.5 & $-$12.5 & 31.44 &  27.77 & 31.12 &  1.41    \\ 
0558$-$504 & 0.137 & 15.18  &  \ldots &  5.4 &  $-$1.3 &  $-$9.5 & 30.44 &  \ldots & 30.25 &  \ldots  \\ 
0624$+$691 & 0.370 & 14.50$^v$  &  \ldots &  7.0 &  $-$0.7 & $-$11.3 & 31.58 &  \ldots & 31.17 &  \ldots  \\ 
0736$+$017 & 0.191 & 16.90  &  235 &  9.4 &  $-$0.9 & $-$11.4 & 30.16 &  26.56 & 29.84 &  1.38    \\ 
0743$-$673 & 1.510 & 16.61  &  \ldots & 11.8 &  0.0 & $-$14.4 & 32.04 &  \ldots & 31.31 &  \ldots  \\ 
0804$+$761 & 0.100 & 15.15  &  1400 &  3.1 &  $-$0.1 & $-$13.0 & 30.11 &  26.78 & 30.07 &  1.28    \\ 
0837$-$120 & 0.198 & 15.78  &  1200 &  5.9 &  $-$1.0 & $-$10.9 & 30.52 &  27.30 & 30.12 &  1.24    \\ 
0859$-$140 & 1.327 & 16.79  &  129 &  5.7 &  $-$0.4 & $-$13.2 & 31.68 &  27.97 & 31.13 &  1.42  \\ 
0906$+$484 & 0.118 & 16.46  &   38 &  1.8 &  $-$1.4 &  $-$9.6 & 29.72 &  25.36 & 29.60 &  1.68    \\ 
0953$+$414 & 0.239 & 14.80$^v$  &  \ldots &  1.3 &  $-$1.5 &  $-$8.6 & 30.97 &  \ldots & 30.81 &  \ldots  \\ 
1001$+$291 & 0.329 & 15.62  &  \ldots &  1.9 &  $-$1.6 &  $-$8.8 & 30.90 &  \ldots & 30.81 &  \ldots  \\ 
1004$+$130 & 0.241 & 15.93  &  $<$  26 &  3.7 &  $-$0.5 & $-$12.5 & 30.55 &  $<$ 25.81 & 30.23 &  $>$ 1.82    \\ 
1007$+$417 & 0.611 & 16.80$^v$  &  \ldots &  1.2 &  $-$1.1 & $-$10.5 & 30.93 &  \ldots & 31.15 &  \ldots  \\ 
1008$+$133 & 1.287 & 16.24  &  $<$  37 &  3.8 &  $-$0.4 & $-$12.9 & 31.81 &  $<$ 27.40 & 31.41 &  $>$ 1.69    \\ 
1011$-$282 & 0.253 & 16.80  &  230 &  5.7 &  $-$1.3 & $-$10.2 & 30.31 &  26.80 & 30.14 &  1.35    \\ 
1012$+$008 & 0.185 & 16.30$^v$  &  \ldots &  3.5 &  $-$1.2 & $-$10.5 & 30.18 &  \ldots & 29.82  &  \ldots \\ 
1048$-$090 & 0.344 & 16.00  &  440 &  3.2 &  $-$1.3 & $-$10.1 & 30.80 &  27.35 & 30.40 &  1.33    \\ 
1049$-$005 & 0.357 & 16.25$^v$  &  \ldots &  3.9 &  $-$0.5 & $-$12.6 & 30.75 &  \ldots & 30.33  &  \ldots \\ 
1100$+$772 & 0.311 & 15.86  &  580 &  3.0 &  $-$1.1 & $-$10.2 & 30.77 &  27.38 & 30.70 &  1.30    \\ 
1103$-$006 & 0.426 & 16.39  &  \ldots &  4.1 &  $-$1.0 & $-$11.0 & 30.84 &  \ldots & 30.61 &  \ldots  \\ 
1104$+$167 & 0.634 & 15.91  &  \ldots &  1.5 &  $-$1.0 & $-$11.1 & 31.32 &  \ldots & 30.96 &  \ldots  \\ 
1116$+$215 & 0.177 & 15.17  &  710 &  1.3 &  $-$1.0 & $-$10.1 & 30.57 &  26.98 & 30.72 &  1.38    \\ 
1127$-$145 & 1.187 & 17.17  &  150 &  4.1 &  $-$1.4 & $-$10.3 & 31.38 &  27.94 & 31.24 &  1.32    \\ 
1136$-$135 & 0.557 & 16.15  &  \ldots &  3.5 &  $-$1.6 &  $-$9.0 & 31.14 &  \ldots & 31.19 &  \ldots  \\ 
1137$+$660 & 0.646 & 16.50  &  335 &  1.0 &  $-$1.0 & $-$10.8 & 31.10 &  27.77 & 31.19 &  1.28   \\ 
1146$-$037 & 0.341 & 16.96  &  530 &  2.5 &  $-$0.8 & $-$12.4 & 30.40 &  27.42 & 29.64 &  1.14    \\ 
1148$+$549 & 0.969 & 16.12$^v$  &  \ldots &  1.2 &  $-$1.0 & $-$10.9 & 31.60 &  \ldots & 31.62  &  \ldots \\ 
1151$+$117 & 0.176 & 15.81$^v$  &  \ldots &  2.5 &  $-$0.8 & $-$11.5 & 30.31 &  \ldots & 30.07  &  \ldots \\ 
1156$+$295 & 0.729 & 14.80  &  \ldots &  1.6 &  $-$0.4 & $-$12.9 & 31.88 &  \ldots & 31.05  &  \ldots \\ 
1156$+$631 & 0.594 & 17.10$^v$  &  \ldots &  1.9 &  $-$1.0 & $-$11.0 & 30.79 &  \ldots & 30.81 &  \ldots  \\ 
1202$+$281 & 0.165 & 15.02  &  870 &  1.7 &  $-$0.9 & $-$11.2 & 30.58 &  27.01 & 29.83 &  1.37   \\ 
1206$+$459 & 1.158 & 16.09$^v$  &  \ldots &  1.3 &  0.1 & $-$14.9 & 31.75 &  \ldots & 30.78 &  \ldots  \\ 
1216$+$069 & 0.334 & 15.68  &  300 &  1.6 &  $-$1.1 & $-$10.6 & 30.89 &  27.15 & 30.62 &  1.43    \\ 
1241$+$176 & 1.273 & 15.38  &   55 &  1.8 &  0.2 & $-$15.3 & 32.11 &  27.56 & 30.77 &  1.75    \\ 
1248$+$401 & 1.030 & 16.36$^v$  &  \ldots &  1.4 &  $-$0.5 & $-$12.7 & 31.56 &  \ldots & 31.10 &  \ldots  \\ 
1259$+$593 & 0.472 & 15.90$^v$  &  \ldots &  1.4 &  $-$0.9 & $-$11.0 & 31.08 &  \ldots & 30.96 &  \ldots  \\ 
1302$-$102 & 0.286 & 15.09  &  540 &  3.4 &  $-$0.8 & $-$11.1 & 31.02 &  27.28 & 30.78 &  1.44   \\ 
1307$+$085 & 0.155 & 15.28  &  840 &  2.1 &  $-$1.1 & $-$10.1 & 30.42 &  26.94 & 30.17 &  1.34    \\ 
1309$+$355 & 0.184 & 15.75$^v$  &  \ldots &  1.0 &  $-$1.2 & $-$10.3 & 30.37 &  \ldots & 29.93  &  \ldots \\ 
\hline
\end{tabular}

\clearpage
\begin{tabular}{lclrrrrrrrr} 
\multicolumn{11}{c}{ TABLE 4 - {\em Continued}} \\
\hline\hline  
\multicolumn{11}{c}{} \\
Name & $z$ &  $B$ & $F_x^{~a}$ & $N_H^{~b}$ & \auv$^{~c}$ &
$b_{uv}^{~d}$  & log~$l_{opt}^{~e}$ & log~$l_{x}^{~f}$ &
log~$l_{uv}^{~g}$ & \aox  \\  
\hline
1317$+$277 & 1.022 & 16.12  &  \ldots &  1.2 &  $-$0.4 & $-$12.9 & 31.65 &  \ldots & 31.39 &  \ldots  \\ 
1318$-$113 & 2.308 & 17.81  &  \ldots &  2.8 &  $-$1.7 &  $-$9.5 & 31.64 &  \ldots & 31.98 &  \ldots  \\ 
1333$+$176 & 0.554 & 15.64  &   49 &  1.8 &  $-$0.9 & $-$11.3 & 31.31 &  26.80 & 31.02 &  1.73    \\ 
1334$+$246 & 0.107 & 15.30$^v$  &  \ldots &  1.1 &  1.1 & $-$18.2 & 30.10 &  \ldots & 28.79 &  \ldots \\ 
1338$+$416 & 1.219 & 16.38$^v$  &  \ldots &  0.9 &  0.0 & $-$14.5 & 31.68 &  \ldots & 30.88 &  \ldots \\ 
1352$+$183 & 0.152 & 15.71  &  530 &  2.0 &  $-$1.4 &  $-$9.5 & 30.23 &  26.72 & 29.90 &  1.35    \\ 
1354$+$195 & 0.720 & 16.48  &  \ldots &  2.2 &  $-$0.4 & $-$13.1 & 31.20 &  \ldots & 30.82 &  \ldots \\ 
1355$-$416 & 0.313 & 15.76  &  620 &  6.2 &  $-$0.5 & $-$12.3 & 30.92 &  27.41 & 30.54 &  1.35    \\ 
1356$+$581 & 1.371 & 17.32  &  \ldots &  1.4 &  $-$0.8 & $-$12.4 & 31.40 &  \ldots & 31.05 &  \ldots \\ 
1415$+$451 & 0.114 & 16.04$^v$  &  \ldots &  1.1 &  $-$1.0 & $-$11.0 & 29.86 &  \ldots & 29.44 &  \ldots \\ 
1425$+$267 & 0.362 & 15.67  &   73 &  1.7 &  $-$0.4 & $-$13.1 & 30.96 &  26.61 & 30.16 &  1.67    \\ 
1444$+$407 & 0.267 & 16.25$^v$  &  \ldots &  1.3 &  $-$1.0 & $-$10.8 & 30.48 &  \ldots & 30.38 &  \ldots \\ 
1512$+$370 & 0.371 & 15.48  &  350 &  1.4 &  $-$1.4 &  $-$9.6 & 31.05 &  27.31 & 30.75 &  1.44    \\ 
1522$+$101 & 1.321 & 16.04$^v$  &  \ldots &  2.9 &  0.1 & $-$14.4 & 31.88 &  \ldots & 31.31 &  \ldots \\ 
1525$+$227 & 0.253 & 16.79  &   38 &  4.2 &  $-$0.6 & $-$12.2 & 30.27 &  26.02 & 30.16 &  1.63    \\ 
1526$+$285 & 0.450 & 16.00$^v$  &  \ldots &  2.6 &  $-$1.2 & $-$10.5 & 31.00 &  \ldots & 30.57 &  \ldots \\ 
1538$+$477 & 0.770 & 16.31$^v$  &  \ldots &  1.6 &  $-$0.1 & $-$13.8 & 31.33 &  \ldots & 30.88 &  \ldots \\ 
1545$+$210 & 0.264 & 16.05  &  780 &  4.3 &  $-$1.2 & $-$10.3 & 30.60 &  27.37 & 30.36 &  1.24    \\ 
1552$+$085 & 0.119 & 16.02  &   20 &  3.4 &  $-$0.1 & $-$13.8 & 29.92 &  25.08 & 29.41 &  1.86    \\ 
1612$+$261 & 0.131 & 16.00  &  710 &  4.1 &  $-$1.0 & $-$11.2 & 30.03 &  26.72 & 29.53 &  1.27   \\ 
1617$+$175 & 0.114 & 15.53  &  285 &  4.1 &  $-$1.4 &  $-$9.4 & 30.10 &  26.20 & 29.93 &  1.50    \\ 
1630$+$377 & 1.471 & 16.26$^v$  &  \ldots &  1.1 &  0.0 & $-$14.3 & 31.87 &  \ldots & 31.30 &  \ldots \\ 
1634$+$706 & 1.334 & 14.90  &   93 &  4.6 &  $-$0.2 & $-$12.9 & 32.40 &  27.83 & 31.92 &  1.75    \\ 
1641$+$399 & 0.595 & 18.70  &  527 &  1.0 &  $-$0.2 & $-$13.4 & 30.15 &  27.90 & 30.84 &  0.86    \\ 
1656$+$053 & 0.887 & 17.00  &  360 &  6.1 &  $-$0.8 & $-$11.6 & 31.28 &  28.07 & 31.21 &  1.23   \\ 
1718$+$481 & 1.084 & 15.63$^v$  &  \ldots &  2.3 &  $-$0.3 & $-$12.9 & 31.89 &  \ldots & 31.74 &  \ldots \\ 
1721$+$343 & 0.206 & 16.80$^v$  &  1700 &  3.0 &  $-$1.4 &  $-$9.2 & 30.06 &  27.49 & 30.51 &  0.99    \\ 
1803$+$676 & 0.136 & 16.04  &  330 &  4.7 &  $-$0.9 & $-$11.1 & 30.07 &  26.42 & 29.91 &  1.40    \\ 
1850$-$782 & 0.162 & 15.80$^v$  &  \ldots &  9.0 &  0.1 & $-$14.6 & 30.45 &  \ldots & 29.50 &  \ldots \\ 
1928$+$738 & 0.302 & 16.80$^v$  &  \ldots &  7.8 &  $-$1.2 & $-$10.1 & 30.52 &  \ldots & 30.57 &  \ldots \\ 
2112$+$059 & 0.466 & 18.90  &  $<$  61 &  6.5 &  $-$0.9 & $-$11.3 & 29.99 &  $<$ 26.75 & 30.82 &  $>$ 1.24   \\ 
2141$+$175 & 0.213 & 15.91  &  165 &  8.2 &  $-$1.4 &  $-$9.4 & 30.61 &  26.51 & 30.50 &  1.57    \\ 
2201$+$315 & 0.297 & 15.56  &  535 &  9.0 &  $-$1.4 &  $-$9.2 & 31.04 &  27.30 & 30.98 &  1.44   \\ 
2251$+$113 & 0.323 & 16.25  &  $<$  63 &  5.1 &  $-$1.2 & $-$10.2 & 30.71 &  $<$ 26.44 & 30.45 &  $>$ 1.64   \\ 
2251$+$158 & 0.859 & 16.57  &  475 &  6.4 &  $-$0.6 & $-$12.3 & 31.44 &  28.17 & 30.99 &  1.26    \\ 
2302$+$029 & 1.044 & 16.33$^v$  &  \ldots &  5.2 &  $-$0.3 & $-$13.2 & 31.66 &  \ldots & 31.26 &  \ldots \\ 
2344$+$092 & 0.672 & 16.08  &  180 &  5.8 &  $-$1.2 & $-$10.5 & 31.39 &  27.53 & 31.02 &  1.48    \\ 
\hline
\end{tabular}\\[0.5ex]

\parbox[b]{7in}{
\hskip0.8in $^a$~Flux in 0.16$-$3.5keV bandpass, in \fcgs~ from Wilkes et al.
(1994), assuming $\ax=-0.5$.

\hskip0.8in $^b$~Galactic column density in $10^{20}$~cm\mtwo.

\hskip0.8in $^c$~UV spectral slope ($f_{\nu}\sim\nu^{\auv}$)

\hskip0.8in $^d$~UV normalization (see text).

\hskip0.8in $^e$~Log rest-frame luminosity in \fnucgs, at 2500\AA.

\hskip0.8in $^f$~Log rest-frame luminosity in \fnucgs, at 2~keV.

\hskip0.8in $^g$~Log rest-frame luminosity in \fnucgs, at 1450\AA.


}

\end{centering}
\clearpage

\begin{centering}
\begin{tabular}{llcccccl} 
\multicolumn{8}{c}{ TABLE 5 } \\
\multicolumn{8}{c}{ UNIVARIATE DISTRIBUTIONS} \\
\hline\hline  
\multicolumn{8}{c}{ } \\
Parameter & \multicolumn{2}{c}{ Sample }  & Median  &  Mean & RMS & Trunc? & N$_{del}^{~b}$ \\ 
          & N$_{total}$ & N$_{limits}^a$  &      &      &       &        &  \\
\hline
\multicolumn{8}{c}{} \\
\multicolumn{8}{c}{Continuum Parameters} \\
\loglopt &  85 & 0 &  30.87 & 30.91 & 0.07 & N & \\
\logluv &  85 & 0 &  30.71& 30.62 & 0.07 & N & \\
\loglx & 49 & 5 &  27.07 & 27.02 & 0.11 & N & \\
\auv &  85 & 0 &  $-$1.09 & $-$1.19 & 0.06  & N & \\
\aox &  49 & 5 &   ~1.38 & ~1.43 & 0.03 & N & \\
\multicolumn{8}{c}{} \\
\multicolumn{8}{c}{Equivalent Widths W$_{\lambda}$ (\AA)} \\
\ovi &   57 & 29 &      6.9 &   5.9 &   0.9 &   1.5 & \\ 
\lya &   77 & 5 &     57.5 &  52.7 &   3.5 &   N & \\ 
\siiv &   71 & 39 &      5.4 &   6.4 &   1.0 &   1.5 & \\ 
\civ  &   63 & 12 &     52.5 &  49.8 &   5.1 &   5.7 & \\ 
\heii &   58 & 31 &     9.9 &  7.9 &   1.6 &   1.2 & \\ 
\ciii &   52 & 25 &     17.8 &  16.5 &   1.9 &   5.2 & \\ 
\multicolumn{8}{c}{} \\
\multicolumn{8}{c}{Ratios of Equivalent Widths} \\
\rovi &   50 & 24 &      0.12 &   0.12 &   0.02 &   0.03 & \\ 
\roviciv &  28 & 12 &    0.16 &   0.16 &   0.02 &   0.08 & 3 \\ 
\rsiiv &   67 & 37 &     0.10 &   0.11 &   0.02 &   0.02 & 2 \\ 
\rciv &   59 & 10 &      0.81 &   0.84 &   0.06 &  N & 2 \\ 
\rheii &   54 & 27 &     0.16 &  0.13 &   0.02 &   0.02 & 3 \\ 
\rciii &   47 & 23 &     0.26 &   0.27 &   0.03 &   0.08 & 2 \\ 
\rciiiciv &   41 & 21 &  0.21 &   0.24 &   0.03 &   0.09 & 6 \\ 
\multicolumn{8}{c}{} \\
\multicolumn{8}{c}{FWHM (km~s\mone) } \\
\ovi & 57 & 29 &   4570 & 4210 & 290 &  2250  \\
\lya & 77 & 5 &  7590 &  7690 & 250 & 3140  \\
\siiv & 70 & 39 &   4790 & 4050 & 550 & 1240  \\
\civ & 63 & 12 &  7940 &  7820 & 360  & N  \\
\ciii & 52 & 25 &  5750 & 6340 & 520 & N \\
\hline
\end{tabular}\\[0.5ex]

\parbox[b]{7in}{
\hskip1.4in $^a$~All limits are upper limits except for \aox.

\hskip1.4in $^b$~Number of lower limits deleted from sample
to allow analysis. 
}

\end{centering}
\clearpage

\label{tbiv}
\begin{centering}
\begin{tabular}{llccccccrrrcc} 
\multicolumn{13}{c}{ TABLE 6 } \\
\multicolumn{13}{c}{ BIVARIATE RESULTS$^{~a}$} \\
\hline\hline  
\multicolumn{13}{c}{ } \\
\multicolumn{2}{c}{ Parameters } & N$_{total}$ & 
\multicolumn{5}{c}{ N$_{limits}$ } &   $P^{~b}$ &  \multicolumn{2}{c}{~ ~
 ~Regression$^{~c}$} & \multicolumn{2}{c}{N$_{changed}^{~d}$} \\ 
 ~ ~~Y  & ~ ~X  &   &  \multicolumn{2}{c}{ X } &
\multicolumn{2}{c}{Y}  & Both & (\%) & Slope~~~ & Intercept  & X  & Y  \\
 &   &   & ~low & up & ~low & up & & &  &  &   &   \\
\hline
\multicolumn{13}{c}{  } \\
 ~\wlya & \luv & 77 & 0 &  0 & 0 &  5 & 0 & $.01$ & $-0.11\pm0.05$ & $5.2\pm1.6$ & 0 & 1  \\
 ~\wlya & \auv & 77 & 0 & 0 & 0 &  5 & 0 & $1.3$ & $0.12\pm0.06$ & $1.9\pm0.1$ & 0 & 1  \\
 ~\wlya & \aox & 45 & 5 & 0 & 0 &  0 & 0 & $.17$ & $-0.38\pm0.23$ & $2.3\pm0.3$ & 1 & 0  \\
 ~FWHM(\lya) & \luv & 77 & 0 & 0 & 0 &  5 & 0 & $1.4$ & $0.06\pm0.01$ & $2.1\pm0.4$ & 0 & 1  \\
 ~FWHM(\lya) & \auv & 77 & 0 & 0 & 0 &  5 & 0 & $.01$ & $-0.09\pm0.03$ & $3.8\pm.04$ & 0 & 0  \\
\multicolumn{13}{c}{  } \\
 ~\wciv & \luv & 63 & 0 &  0 & 0 & 12 & 0 & $.02$ & $-0.22\pm0.05$ & $8.3\pm1.6$ & 0 & 3  \\
\multicolumn{13}{c}{  } \\
 ~\wciii & \luv & 52 & 0 &  0 & 0 & 25 & 0 & $0.4$ & $-0.27\pm0.05$ & $9.6\pm1.5$ & 0 & 9  \\
 ~\wciii & \lx & 36 & 0 & 2 & 0 & 15 & 0 & $0.8$ & $-0.19\pm0.06$ & $6.5\pm1.6$ & 0 & 6  \\
 ~\rciii & \lx & 33 & 0 & 2 & 0 & 15 & 1 & $.03$ & $-0.23\pm0.10$ & $5.6\pm2.9$ & 1 & 5  \\
\multicolumn{13}{c}{  } \\
 ~\wheii & \luv & 58 & 0 &  0 & 0 & 31 & 0 & $.01$ & $-0.30\pm0.07$ & $9.9\pm2.1$ & 0 & 8  \\
 ~\wheii & \lx & 37 & 0 &  3 & 0 & 17 & 0 & $.5$ & $-0.27\pm0.15$ & $8.1\pm4.1$ & 0 & 5  \\
 ~\rheii & \luv & 54 & 0 &  0 & 0 & 27 & 0 & $.7$ & $-0.21\pm0.08$ & $5.5\pm2.5$ & 0 & 5  \\
 ~\rheii & \lx & 35 & 0 &  3 & 0 & 15 & 0 & $.7$ & $-0.24\pm0.31$ & $5.7\pm8.4$ & 0 & 5  \\
\hline
\end{tabular}\\[0.5ex]

\hskip0.7in {\parbox[b]{6in}{
$^a$~Includes only parameter pairs with data for $N>30$
QSOs and significant ($P<2\%$) correlations in {\em both} generalized
Spearman and Kendall's rank tests.  

$^b$~Probability of achieving the Spearman Rank correlation assuming the null
hypothesis (no correlation).

$^c$~Fits are from Schmitt's 2-D Kaplan-Meier 
regression to the logarithm of listed parameters (except for
spectral slopes \auv~ and \aox).

$^d$~Outlying limits changed to detections at bin edges.


}}
\hoffset0in
\end{centering}